\documentclass[10pt,twocolumn,twoside,journal]{IEEEtran}

\usepackage[nocomma]{optidef}
\usepackage{braket}
\usepackage{booktabs}
\usepackage{mathtools} 
\DeclarePairedDelimiter{\ceil}{\lceil}{\rceil}
\usepackage{csquotes}
\usepackage{makecell}
\usepackage{flushend}

\usepackage{epsfig} 
\usepackage{enumerate}
\usepackage{url}
\usepackage{times} 
\usepackage{amsmath} 
\usepackage{amssymb}  
\usepackage{psfrag}
\usepackage{graphicx,color}
\usepackage{cleveref}
\usepackage{algorithm}
\usepackage{algpseudocode}
\usepackage{amsthm}  
\usepackage[footnotesize]{subfigure}
\usepackage[format=default,footnotesize]{caption}
\usepackage{multirow}  
\usepackage{colortbl} 
\usepackage[table]{xcolor} 
\usepackage{circuitikz}
\usepackage{import}
\usepackage{enumitem}

\graphicspath{figures/}

\newcommand{\eg}{\mbox{e.g.,}}
\newcommand{\ie}{\mbox{i.e.,}}
\newcommand{\etc}{\mbox{etc.}}

\crefname{equation}{}{}

\renewcommand{\arraystretch}{1.2}   
\newlength{\figWidth}
\newlength{\figHeight}

\newlength{\figSpace} 
\newlength{\tableSpace} 

\usepackage{cite}

\definecolor{cyan}{rgb}{0.0, 1.0, 1.0}

\makeatletter

\def\thickhline{%
  \noalign{\ifnum0=`}\fi\hrule \@height \thickarrayrulewidth \futurelet
   \reserved@a\@xthickhline}
\def\@xthickhline{\ifx\reserved@a\thickhline
               \vskip\doublerulesep
               \vskip-\thickarrayrulewidth
             \fi
      \ifnum0=`{\fi}}
\makeatother

\newlength{\thickarrayrulewidth}
\begin{document}

\title{Quantum Computing in the Computational Landscape of Power Electronics: Vision and Reality}

\author{Nikolaos G. Paterakis, \IEEEmembership{Senior Member, IEEE}, Petros Karamanakos, \IEEEmembership{Senior Member, IEEE}, \\ Corey O'Meara, and Georgios Papafotiou, \IEEEmembership{Member, IEEE} 
\thanks{Nikolaos G. Paterakis and Georgios Papafotiou are with the Department of Electrical Engineering, Eindhoven University of Technology, 5600 MB, Eindhoven, The Netherlands (e-mail: n.paterakis@tue.nl; g.papafotiou@tue.nl).}
\thanks{Petros Karamanakos is with the Faculty of Information Technology and Communication Sciences, Tampere University, 33101 Tampere, Finland (e-mail: p.karamanakos@ieee.org).}
\thanks{Corey O'Meara is with E.ON Digital Technology GmbH, 30457, Hanover, Germany (e-mail: corey.o'meara@eon.com).}
}

\maketitle

\thispagestyle{empty}

\begin{abstract}


Quantum computing is rapidly emerging as a promising technology for solving complex optimization problems that arise in various engineering fields. Therefore, it holds significant promise to transform the computational foundations of power electronics. Motivated by this potential, this paper adopts a visionary perspective to examine how quantum computing could influence the evolution of power electronics in areas such as converter design, control, modulation, simulation workflows, and beyond. Within this framework, the current status, limitations, and anticipated progress of quantum algorithms and hardware are discussed, together with their potential to enable efficient solutions to large-scale, multiobjective, mixed-integer optimization problems. To place these developments in context, the paper begins with a concise tutorial on fundamental concepts in quantum computing, serving as both an introduction to the field and a bridge to its potential applications in power electronics. As a first step in this direction, the use of quantum computing for solving offline mixed-integer optimization problems commonly encountered in power electronics is examined. To this end, a simplified power electronics design problem is reformulated as a quadratic unconstrained binary optimization (QUBO) problem and executed on quantum hardware, despite current limitations such as low qubit counts and hardware noise. This demonstration marks a pioneering step towards leveraging quantum computing in power electronics and motivates the value of early adoption and exploration. Building on these insights, the paper outlines a forward-looking vision in which quantum computing becomes an integral part of the computational landscape of power electronics, guiding its transition from classical to quantum-enabled design and operation.

\end{abstract}


\section{Introduction}
\label{sec::Intro}

\IEEEPARstart{R}{ichard} Feynman has been famously quoted saying that \enquote{I think I can safely say that nobody understands quantum mechanics}. Although the authors cannot claim comparable scientific authority, we do believe that a parallel argument can be made for quantum computing, at least when it pertains to the field of power electronics. Despite the fact that quantum computing carries the reputation of a paradigm-altering technology due to its potential to offer unprecedented computational performance, the translation of its potential into tangible benefits for power electronics research and applications remains, at best, obscure, at least for the practicing power electronics engineer. 


To better understand the potential role and impact of quantum computing in power electronics, it is first necessary to examine how computation is currently used in the field. Power electronics, as a multidisciplinary discipline, requires its practitioners to be all-around players. Computation has traditionally played a secondary role, but it is now rapidly reshaping the field~\cite{GWH^+23, LY23, CZKR23, FHBB24, LZ24, SZKM24}. This shift is evident in both online and offline domains. In the former area, powerful microprocessors and field-programmable gate arrays (FPGAs) enable the real-time implementation of increasingly sophisticated and computationally demanding control algorithms~\cite{GLF^+17, MAA18, SDHB18, WGL^+19, MSK^+19}. In the offline computational world, advanced tools and methods facilitate the exploration of vast design spaces~\cite{RNG^+09, PKD^+17, PKD^+20, PGKK20}, while accurate simulation models help narrow the gap between theoretical designs and experimental validation~\cite{GTF^+20, DD23}.

Despite these advances, certain limitations remain, stemming from fundamental challenges associated with the mathematical nature of the underlying problems. More specifically, many control problems in power electronics involve the selection of suitable switch positions among a number of discrete choices due to the switching nature of power converters~\cite{KGK16, KG20, ZVG^+23, ZVG^+24}. In a parallel fashion, the design phase of power electronic systems entails solving complex combinatorial optimization problems, wherein designers are required to select elements out of discrete sets, including, but not limited to, converter topologies, output voltage levels, magnetic materials, switches, and type of cooling. The way these problems are approached and solved has a significant impact on the final design and subsequent performance of the system.

Although the two worlds, \ie{} online control and offline design, may appear to be far apart in a first reading, a closer inspection reveals a common underlying mathematical foundation. Specifically, both problem classes involve the exploration of optimal solutions within discrete sets, thus falling under a class of mathematical problems known as combinatorial optimization problems. This class of problems is also encountered in other scientific and technological fields and is notoriously affected by the \emph{curse of dimensionality}, where the computational complexity increases exponentially with the size of the problem. Although existing techniques for these problems, such as branch-and-bound, cutting planes, \etc{}, can be highly effective in practice, they do not provide guarantees on the maximum execution time. In the worst case, the theoretical computational burden remains equivalent to that of the brute-force approach of exhaustive enumeration. This is a fundamental limitation of the existing computation technologies and cannot be circumvented within the classical computing paradigm.

This is where quantum computing offers the potential for a computational advantage. We are currently in the so-called noisy intermediate-scale quantum (NISQ) era~\cite{Preskill_2018}, characterized by quantum processors with a limited number of physical qubits and imperfect qubit control. Despite these limitations, quantum computing has attracted significant attention from researchers aiming to achieve quantum utility, \ie{} the ability to efficiently solve practically relevant problems that are hard or intractable for classical approaches~\cite{McGeoch_2022, King_2025}. As a result, in addition to fundamental research, a growing body of literature is emerging on quantum algorithm applications in various domains, such as logistics~\cite{phillipson_2025}, finance~\cite{Buonaiuto_2023} and electrical power systems~\cite{Morstyn_2024}. 

There are several reasons why quantum computing is believed to be advantageous for optimization. First, many combinatorial optimization problems have a natural mapping to quantum systems, through their reformulation to Ising models or quadratic unconstrained binary optimization (QUBO) problems~\cite{Lucas_2014}. Second, despite the aforementioned exponential complexity of these problems, there are quantum algorithms---such as Grover's search---that can offer a theoretical quadratic speedup over classical brute-force search~\cite{Abbas_2024}.
These insights suggest that quantum optimization algorithms may be advantageous, at least for some problem classes. In this context, it is possible to devise hybrid quantum-classical optimization algorithms, which allow quantum computers to be used as coprocessors alongside classical computing resources~\cite{Gambella_2020, Paterakis_2023}.\looseness=-1
 
It is therefore this potential that makes quantum computing particularly intriguing for power electronics. However, the current maturity of the technology, and its limited accessibility and availability to the average power electronics designer and researcher, render its immediate application questionable. This is further compounded by that fact that the \enquote{cookbook} of power electronics design currently does not contain any recipes for using quantum computing. There is limited awareness of the types of problems that quantum computing can address, and how these problems might be formulated within a quantum framework. This lack of clarity and insight not only obscures the actual improvements that quantum computing might bring, but also, and perhaps more crucially, the ways in which it could be meaningfully implemented for solving problems in power electronics. 

Motivated by these observations, this paper takes a visionary perspective on the intersection of quantum computing and power electronics. Its aim is twofold. First, it provides a concise, tutorial-style introduction to the fundamental principles of quantum computing and quantum optimization, bridging conceptual understanding with practical formulation. Second, it outlines a forward-looking vision of how this emerging technology could reshape the computational landscape of power electronics in the coming decades. By examining both the current maturity of quantum computing from a user perspective and its anticipated trajectory, the paper aims to assess its potential to deliver practically relevant results in the near future and to stimulate early awareness, informed discussion, and creative exploration within the power electronics community.


To this end, we begin in Section~\ref{sec::Tut} with a concise yet comprehensive introduction to the basic notions and concepts of quantum computing that are essential for entering the field. Following, in Section~\ref{sec::QuantPE}, we present a simplified design problem at the level of an undergraduate power electronics course---namely, the filter design of a dc-dc boost converter---and show how it can be formulated as a combinatorial optimization problem and subsequently cast as a QUBO problem. 
The example is intentionally kept as simple as possible; it does not fully reflect the complexity of real-world power converter design, but it serves as a clear and useful case study for exploring the intricacies and current limitations of applying quantum optimization algorithms. The resulting QUBO problem is subsequently solved in Section~\ref{sec::Results}, where detailed numerical and experimental results---in the context of quantum computing, \ie{} the formulated QUBO problem is executed on quantum hardware---are presented. To the best of the authors' knowledge, this represents the first instance of a power electronics-related problem---however elementary---being solved on a quantum computer, marking a noteworthy (symbolic) milestone for the field. In a next step, Section~\ref{sec::Vision} discusses the current limitations of quantum technology and presents a forward-looking perspective on how anticipated advancements could transform the computational landscape of power electronics. Considering short-, mid-, and long-term horizons, it also highlights which problems may become tractable and explores potential implications for design, control, simulation, and other emerging applications. Finally, Section~\ref{sec::Concl} summarizes our findings and conclusions.

\section{Quantum Optimization}
\label{sec::Tut}


The foundational idea behind quantum computing is to leverage quantum mechanical principles to solve computational problems more efficiently than classical methods in certain cases. Such capabilities make it particularly appealing for solving challenging optimization problems. Among the various paradigms, the most prominent are \emph{gate-based quantum computing} and \emph{adiabatic quantum computing}~\cite{mcgeogh_2020}. The latter has inspired a practical heuristic approach known as quantum annealing (QA), with practical systems developed by companies such as D-Wave~\cite{Dwave_website}. QA is tailored specifically for solving certain classes of combinatorial optimization problems, whereas the gate-based model is suitable for universal quantum computation. \looseness=-1

In the gate-based paradigm, quantum algorithms are consructed by applying a sequence of quantum gates, which are physically realizable (mathematical) operations that manipulate the state of quantum bits (qubits) in a controlled way. These gates are described by unitary transformations, which preserve key quantum properties such as superposition and entanglement. After applying these gates, (part of) the quantum system is measured, and the outcome---\ie{} the resulting quantum state---reveals information about the solution to the computational problem of interest. Gate-based quantum computing is actively being pursued by a multitude of companies, including IBM \cite{IBM_roadmap} and Google \cite{Google_AI}. 

In this section, we discuss how gate-based computer can be used to solve combinatorial optimization problems. An introductory tutorial on the basic concepts of gate-based quantum computing that is necessary to discuss the application of quantum computing to combinatorial optimization problems, including those arising in power electronics, is provided in Appendix~\ref{app:quantum-tutorial}.


\subsection{Combinatorial Optimization Problems}

\subsubsection{Cost Hamiltonian}
\label{combinatorial_opt_1}

A class of optimization problems that naturally arise in many real-world applications and are widely studied in the context of quantum optimization is the so-called QUBO problems~\cite{glover_2022}. These aim to compute the optimal binary vector $\mathbf{x}^* \in \left\{ 0, 1 \right\}^n$, and are of the form
\begin{mini}
{\mathbf{x} \in \left\{ 0, 1 \right\}^n}
{\mathbf{x}^T Q \mathbf{x} + \mathbf{c}^T \mathbf{x}}
{\label{general_qubo}}{} 
\end{mini}
where $Q \in \mathbb{R}^{n \times n}$ is a symmetric matrix encoding quadratic coefficients, and $\mathbf{c} \in \mathbb{R}^n$ the linear coefficient vector. QUBO problems can be converted into an equivalent Ising spin glass model through the linear transformation $\mathbf{x} = \frac{\mathbf{1}_n - \mathbf{z}}{2}$, where $\mathbf{z} \in \left\{-1, +1 \right\}^n$ and $\mathbf{1}_n$ is the $n$-dimensional vector of ones. This yields the equivalent Ising model formulation
\begin{mini}
{\mathbf{z} \in \left\{ -1, +1 \right\}^n}
{\mathbf{z}^T R \mathbf{z} + \mathbf{h}^T \mathbf{z}}
{\label{general_ising}}{}
\end{mini}
where $R$ and $\mathbf{h}$ are computed in a straightforward manner from $Q$ and $\mathbf{c}$, respectively. 

Problem \eqref{general_ising} is suitable for quantum computation by promoting the decision variables $z_i$ to Pauli Z gates with eigenvalues of $+1$ and $-1$, whereby solving the optimization problem is equivalent to finding the ground state of the cost Hamiltonian $\mathcal{H}^C=\sum_{i=1}^{n}h_i\sigma_i^z + \sum_{i=1}^{n}\sum_{j>i}^{n} r_{ij} \sigma_i^z \sigma_j^z$, where $\sigma_i^z$ is the Pauli Z gate acting on the $i$-th qubit.

\subsubsection{Constraint Handling}
\label{combinatorial_opt_2}

QUBO (and, therefore, Ising) problems are unconstrained by definition. However, practical optimization problems typically contain equality and inequality constraints. Constrained combinatorial optimization problems are compatible with the QUBO framework by allowing the augmentation of the objective function with penalty terms that enforce constraints.

Linear equality constraints of the form $A \mathbf{x} = \mathbf{b}$ can be easily converted into penalty terms of the form $(A \mathbf{x} - \mathbf{b})^2$ that can be incorporated into \eqref{general_qubo}, weighted by an appropriate penalty weight vector $\boldsymbol \lambda$. The choice of the penalty weights is critical, since they must be large enough to enforce constraints but not so large as to dominate the objective function \cite{Garcia_2022}.

General linear inequality constraints of the form $A \mathbf{x} \leq \mathbf{b}$ must first be converted into equality constraints by introducing appropriate slack variables, which are in turn represented by binary variables. For example, consider the inequality constraint $\sum_{i=1}^{n}a_ix_i\leq b$, which is equivalent to $\sum_{i=1}^{n}a_ix_i + s = b$ with $s \in [0, b-\min_\mathbf{x}a_i x_i] \subseteq \mathbb{Z}$. Using $K=\lceil \log_2(\max_{\mathbf{x}}(b-a_i x_i) +1)\rceil $ binary variables $s_k$, the integer slack variable $s$ can be decomposed into binary form $s=\sum_{k=0}^{K-1}2^k s_k$. Alternative encodings can also be used \cite{Tamura_2021, Dominguez_2023}. It should be noted that the approach described above assumes that the coefficients of the constraints are integer valued. If some elements of $A$ are real valued, then they can be converted with arbitrary accuracy into integers by multiplying both sides of the constraint with an appropriate power of 10. \looseness=-1

Introducing slack variables to represent linear inequality constraints allows for an exact encoding within the optimization problem. This, however, comes at the expense of increasing the number of qubits required to represent the optimization problem on a quantum computer, which is challenging for NISQ hardware. The recently proposed unbalanced penalization method \cite{Barrera_2024} provides a way to approximately enforce inequality constraints without introducing any slack variables. The idea stems from the observation that a monotonic function, such as the exponential, could be used to penalize constraint violations while guaranteeing that the penalty diminishes for feasible solutions. However, the exponential function is not compatible with the QUBO formulation, therefore its second-order Taylor expansion is used instead, despite the fact that it is not monotonic.

Finally, objective function terms that involve higher-order interactions or constraints with non-linear terms must be first processed through a quadratization procedure~\cite{Dattani_2019}. This reduces the resulting penalty function to contain at most quadratic terms. 

\subsection{Quantum Approximate Optimization Algorithm}
\label{sec:qaoa_theory}

\begin{figure}[t]
    \centering
    \includegraphics[width=1\linewidth]{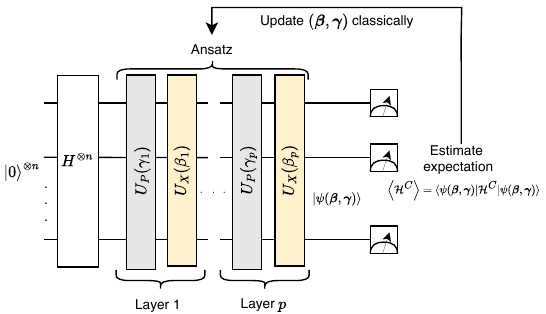}
    \caption{Schematic illustration of QAOA.}
    \label{fig:qaoa}
    \vspace{-6pt}
\end{figure}

Several approaches to designing quantum optimization algorithms have been proposed~\cite{Abbas_2024}. In this paper, we apply one of the most popular hybrid quantum-classical optimization algorithms, namely the quantum approximate optimization algorithm (QAOA)~\cite{farhi_2014} that can be executed on NISQ devices to provide approximate solutions to QUBO problems. 

QAOA belongs to the family of variational quantum algorithms~\cite{Cerezo_2021}. The objective is to determine a parametrized quantum circuit (called ansatz) that solves the problem of interest. To achieve this, in QAOA, the state of the $n$ qubits (each corresponding to a decision variable) is initialized at a uniform superposition $\ket{+}^{\otimes n}$. Then, a sequence of alternating unitary operators $U_X(\beta)=e^{-i\beta \mathcal{H}^M}$ and $U_P(\gamma)=e^{-i\gamma \mathcal{H}^C}$ is applied $p\geq1$ times ($p$ is also called the number of layers). $\mathcal{H}^C$ is the cost Hamiltonian and $\mathcal{H}^M$ is the mixing Hamiltonian. Given an objective function in the form of \eqref{general_ising}, the cost Hamiltonian is found by promoting each variable $z$ to a Pauli Z operator. The commonly used mixing Hamiltonian is $\mathcal{H}^M=\sum_{i=0}^{n} \sigma_i^X$.  Finally, a measurement on the computational basis is performed to evaluate the expected value of the cost Hamiltonian $\bra{\psi(\boldsymbol \beta, \boldsymbol \gamma)}\mathcal{H}^C\ket{\psi(\boldsymbol \beta, \boldsymbol \gamma)}$. The algorithm is hybrid in the sense that the parameters $(\boldsymbol \beta, \boldsymbol \gamma) \in \mathbb{R}^p\times \mathbb{R}^p$ are updated using a classical optimizer so that the expected value of the cost Hamiltonian is minimized. For the optimized set of parameters $(\boldsymbol \beta^*, \boldsymbol \gamma^*)$, the state $\ket{\psi(\boldsymbol \beta^*, \boldsymbol \gamma^*)}$ encodes the solution to the optimization problem. A schematic illustration of QAOA is shown in Fig.~\ref{fig:qaoa}. \looseness=-1

QAOA is considered a promising NISQ algorithm because $U_X$ can be implemented using a single layer of $R_X$ gates. Moreover, operator $U_P$ is realized using $R_{ZZ}$ gates between qubits that correspond to variables involved in quadratic terms of the Ising model, and $R_Z$ gates for linear terms. This structure results in relatively shallow quantum circuits, well suited for NISQ devices. A didactic example of applying QAOA is provided in Appendix~\ref{app:qaoa-example}.

The performance of QAOA depends on the number of layers $p$ and the choice of parameters $\boldsymbol{\beta}$ and $\boldsymbol{\gamma}$. Specifically, for $p \rightarrow \infty$, QAOA would converge to the optimal solutions since it corresponds to adiabatic evolution. For didactic purposes, in this paper we apply the original version of the QAOA algorithm~\cite{farhi_2014}. However, several variants of the QAOA algorithm have been proposed, see, \eg{} \cite{Blekos_2024} and~\cite{bucher_2025}.

\subsection{Considerations for Practical Implementation}

\subsubsection{Workflow}

\begin{figure}[t]
    \centering
    \includegraphics[width=1\linewidth]{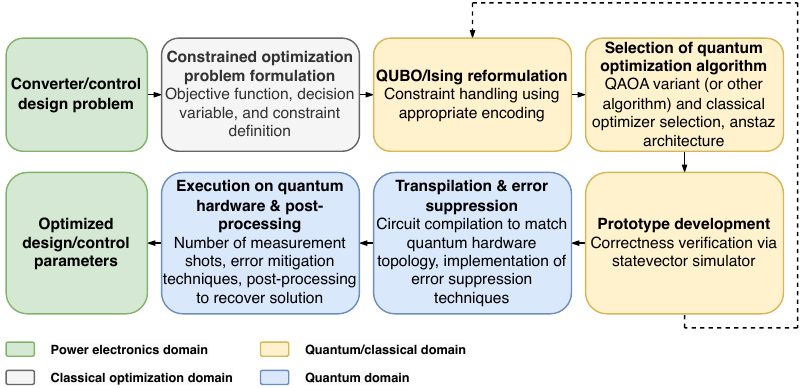}
    \caption{Engineering workflow for implementing a constrained optimization problem of industry relevance on quantum hardware.}
    \label{fig:engineering_workflow}
    \vspace{-6pt}
\end{figure}

Solving constrained optimization problems on gate-based quantum computers is a multi-step process, as illustrated in Fig. \ref{fig:engineering_workflow}.

Given a practical problem of interest, the process begins by formulating a mathematical model with a well-defined objective function and constraints set. As discussed in Section~\ref{combinatorial_opt_1}, the most common class of optimization problems studied in the context of quantum computing is QUBO. Converting a constrained optimization problem into QUBO requires transforming constraints into penalty terms, as discussed in Section~\ref{combinatorial_opt_2}. The trade-off between introducing slack variables and exactness when handling inequality constraints must be carefully considered depending on the range of each constraint. In the case of mixed-integer problems, applying a decomposition-based solution approach may be required~\cite{Gambella_2020, Paterakis_2023}.\looseness=-1

Once the QUBO Hamiltonian has been devised, the next step involves selecting the specific quantum optimization algorithm, such as the QAOA described in Section~\ref{sec:qaoa_theory}. In the case of hybrid variational quantum algorithms, the classical optimizer used to update the parameters of the ansatz and its hyperparameters must also be selected~\cite{Blekos_2024}. The choice of ansatz and initialization strategy affects performance and resource requirements~\cite{Zhou_2020}. Currently, as discussed in Section~\ref{sec:quantum_costs}, quantum processing units (QPU) are primarily accessible via the cloud, which may entail significant execution costs. Therefore, testing a prototype version of the algorithm and tuning parameters using a statevector simulator (either ideal or modeling noise of a particular QPU) that runs on classical computing resources is recommended before execution on real hardware. Simulators are readily available in quantum development frameworks such as PennyLane \cite{pennylane} and \mbox{Qiskit \cite{qiskit}.}

Similar to classical computing, there exist sets of universal quantum gates that can be used to construct any quantum circuit. Moreover, a quantum algorithm may require interactions between qubits that are not directly connected in the physical hardware. Therefore, before executing a quantum algorithm on a QPU, the algorithm must be \emph{transpiled}, \ie{} converted into a form compatible with the instruction set architecture and qubit connectivity topology of the target quantum device~\cite{kremer_2025}. At this stage, various hardware-aware circuit optimization and error suppression techniques, such as dynamical decoupling~\cite{Ezzell_2023}, can be considered. It should be noted that from an implementation perspective, there exist software tools that automate this process, \eg{} Q-CTRL Fire Opal~\cite{QCTRL}.

Finally, the transpiled and optimized circuit is executed on the chosen QPU. In contrast with statevector simulators, execution on real hardware entails preparing and measuring the output of the quantum circuit for a number of trials called \emph{shots} in order to obtain a distribution of candidate solutions. Therefore, it is important to determine the number of shots, considering that more shots increase the likelihood of observing feasible or optimal solutions, but at the same time increase the execution time.

\subsubsection{Quantum Computing Resources and Cost}\label{sec:quantum_costs}

A variety of QPUs based on different hardware technologies are accessible through the cloud from major providers such as IBM Quantum \cite{IBM_roadmap}, Amazon Braket \cite{Amazon_braket}, Microsoft Azure Quantum \cite{Microsoft_Azure}, and Google Quantum AI \cite{Google_AI}. At the time of writing, commercial cloud rates vary considerably across providers and depend on factors such as the chosen backend QPU, the number of samples requested, and whether access is billed on a pay-as-you-go basis or through a commercial contract. Nevertheless, most platforms offer free access tiers that are sufficient for experimentation and proof-of-concept studies. Moreover, both state-vector simulators and emulators of real QPUs can be executed on classical resources, either locally or on the cloud, at negligible cost. For this reason, we believe that the initial cost barrier for researchers interested in exploring quantum computing for real-world applications is relatively low. 

\section{Power Electronics Design Problem}
\label{sec::QuantPE}

In this section, we present a simple yet illustrative design example from power electronics, namely, specifying a basic dc-dc boost converter filter. The design problem is initially formulated as a constrained mixed-integer non-linear program (MINLP) and then systematically transformed into a QUBO problem, making it ameanable to solution by quantum optimization algorithms. This intentionally simplified example serves to clearly demonstrate the process of translating a conventional power electronics design problem into a form compatible with quantum computing, thereby highlighting the key conceptual and technical steps needed to bridge the two domains.\looseness=-1

\subsection{Mixed-Integer Non-Linear Programming Formulation}
\label{sec::DesPE}

\begin{figure}[t!]
    \setlength{\figHeight}{3.cm}
    \setlength{\figWidth}{0.48\textwidth}
    \centering
    \includegraphics[width=\figWidth]{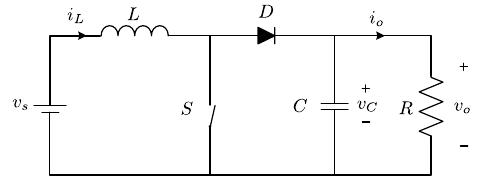}
    \caption{Topology of the dc-dc boost converter.}
    \label{fig::BoostCircuit}
    \vspace{\figSpace}
\end{figure}

Assume a dc-dc boost converter, as shown in Fig.~\ref{fig::BoostCircuit}. The goal is to design its output filter by selecting appropriate values for the inductor $L$ and the capacitor $C$ such that specific performance requirements are met while minimizing the total monetary cost. These passive components are selected from given sets $\mathcal{L} = \{L_1, L_2, \ldots, L_n \}$ and $\mathcal{C} =  \{C_1, C_2, \ldots, C_m \}$, where $n , m \in \mathbb{N}$.

The chosen design criteria  require that the (peak) inductor current ripple $\Delta i_L$, and the (peak) capacitor voltage ripple $\Delta v_C$, remain below predefined maximum values, denoted by $\Delta i_{L,\max}$ and $\Delta v_{C,\max}$, respectively. In addition, the resonance frequency of the output filter $f_{\textrm{res}}$ should be significantly lower than the switching frequency $f_{\textrm{sw}}$ of the converter. These design constraints can be formulated as follows
\begin{subequations} 
\begin{align}
 \Delta i_L &= \frac{v_s}{2 L_i} \frac{d}{f_{\textrm{sw}}} \leq \Delta i_{L,\max} \,,
\label{eq::iLrippl} 
\\
 \Delta v_C &= \frac{v_o}{2 R C_j} \frac{d}{f_{\textrm{sw}}} \leq \Delta v_{C,\max} \,,
\label{eq::vCrippl}
\\
 f_{\textrm{res}} &= \frac{1}{2 \pi \sqrt{L_i C_j}} \leq \frac{f_{\textrm{sw}}}{\kappa} \,,
 \label{eq::fRes}
\end{align}%
\label{eq::rippl}%
\end{subequations}%
where $L_i \in \mathcal{L}$, $i = 1, 2, \ldots, n$, and $C_j \in \mathcal{C}$, $j = 1, 2, \ldots, m$, are the to-be-selected candidate values for the filter inductance and capacitance, respectively. Moreover, $v_s$ and $v_o$ are the input and output voltage levels of the converter, respectively. To keep this educational example straightforward, the converter is assumed to supply a constant resistive load $R$, and operates at a fixed duty cycle $d$ and switching frequency $f_{\textrm{sw}}$.  Finally, $\kappa > 1$ in~\eqref{eq::fRes} is a predefined safety factor used to ensure adequate attenuation of switching harmonics, with $\kappa \geq 10$ typically assumed. All relevant converter parameters and design considerations are summarized in Table~\ref{tab::convParam}.

\begin{table}[!t]
  \footnotesize                                 
  \setlength{\tabcolsep}{6pt}                   
  \caption{Parameters of the dc-dc boost converter example design.}
  \centering
  \begin{tabular}{|c c c|}
    \hline
    \textbf{Parameter} & \textbf{Symbol} & \textbf{Value} \\
    \hline
    Input voltage & $v_s$ & $12\,$V 
    \\
    Resistive load & $R$ & $10\,\Omega$
    \\
    Duty cycle & $d$ & $0.5$ 
    \\
    Switching frequency & $f_{\textrm{sw}}$ & $100\,$kHz
    \\
    Maximum inductor current ripple & $\Delta i_{L,\max}$ & $3\,$A
    \\
    Maximum capacitor voltage ripple & $\Delta v_{C,\max}$ & $0.2\,$V
    \\ 
    Attenuation factor & $\kappa$ & $15$
    \\
    \hline
  \end{tabular}
  \label{tab::convParam}
\end{table}

Given the design criteria and objectives outlined above, the example design task can be formulated as an MINLP. To this end, we define the vector of decision variables $\mathbf{x}=\left[ (\mathbf{x}^{L})^T ~ (\mathbf{x}^{C})^T \right]^T$, where $\mathbf{x}^L$ and $\mathbf{x}^C$ are the vectors of binary variables indicating whether a certain inductor or capacitor is selected. Furthermore, the (monetary) cost of each candidate inductor and capacitor is denoted by $k_i^L$ and $k_j^C$, respectively. Accordingly, the optimization problem takes the following form
\begin{mini!}
{\mathbf{x}}
{\sum_{i \in \mathcal{L}} k_i^L x_i^L+ \sum_{j \in \mathcal{C}} k_j^C x_j^C \label{design_objective}}
{\label{design_problem}}{}
\addConstraint{\sum_{i \in \mathcal{L}} x_i^L}{=1   \label{select_one_L}}{}
\addConstraint{\sum_{j \in \mathcal{C}} x_j^C}{=1   \label{select_one_C}}{}
\addConstraint{\sum_{i \in \mathcal{L}} \sum_{j \in \mathcal{C}} x_i^L x_j^C L_i C_j}{\geq \left( \frac{\kappa}{2 \pi f_{\textrm{sw}}} \right)^{\!2} \label{design_nlc}}{}
\addConstraint{\sum_{i \in \mathcal{L}} x_i^L L_i}{\geq \frac{d \, v_s}{2 f_{\textrm{sw}} \Delta i_{L,\max}} \label{design_L_con}}{}
\addConstraint{\sum_{j \in \mathcal{C}} x_j^C C_j}{\geq \frac{d \, v_o}{2 f_{\textrm{sw}} R \Delta v_{C,\max}} \label{design_C_con}}{} \,.
\end{mini!}
Building on problem~\eqref{design_problem}, the remainder of this section demonstrates how it can be reformulated as a QUBO problem suitable for implementation on a quantum device.

\subsection{Quadratic Unconstrained Binary Optimization Problem Formulation}

As QUBO problems are unconstrained by definition, the formulation of the QUBO version of problem~\eqref{design_problem} begins by incorporating its constraints into the objective function. To this end, the equality constraints~\eqref{select_one_L} and~\eqref{select_one_C}, which ensure that the converter filter includes exactly one inductor and one capacitor, are added to the objective function as quadratic penalty terms, given in~\eqref{design_qubo_1} and~\eqref{design_qubo_2}, respectively.

Handling constraint~\eqref{design_nlc} in the QUBO framework is challenging. Firstly, we observe that~\eqref{design_nlc} includes bi-linear terms, which would result in quartic terms if added to the objective function directly. To avoid this, we introduce auxiliary variables $z_{ij}$ such that $z_{ij}=x_i^L x_j^C, \forall i \in \mathcal{L}, \forall j \in \mathcal{C}$. The new constraint has a known (Rosenberg~\cite{Rosenberg_1975}) quadratization $h_{ij}(z_{ij}, x_i^L, x_j^C)=3z_{ij}+x_i^L x_j^C - 2 x_i^L z_{ij} - 2 x_j^C z_{ij}$ that can be added as a penalty to the objective function, see term~\eqref{design_qubo_3}. Moreover, the penalty term~\eqref{design_qubo_4} is introduced to ensure that exactly one valid inductor--capacitor combination is selected. Secondly, to exactly encode the inequality constraint, it must be reformulated as an equality by introducing slack variables. Although several encoding strategies exist for this transformation (see, \eg{} \cite{Tamura_2021}), introducing slack variables can significantly enlarge the solution space, compromising solution quality, and potentially exceeding the capabilities of current NISQ hardware even for small problem instances. For example, due to the wide range of inductance and capacitance values, in our experiments, at least 11 slack variables would be required to exactly encode \eqref{design_nlc} using binary encoding. To avoid these issues, we adopt the recently proposed unbalanced penalization approach~\cite{Barrera_2024}, which results in the additional terms given in~\eqref{design_qubo_5} and~\eqref{design_qubo_6}.

Finally, constraints~\eqref{design_L_con} and~\eqref{design_C_con} could, in principle, be handled similarly to~\eqref{design_nlc}. However, we observe that problem~\eqref{design_problem} can be simplified by removing these constraints and preprocessing the sets of inductors and capacitors to include only those components that satisfy the current and voltage ripple requirements. This procedure can be done in at most $\mathcal{O}(\vert \mathcal{L} \vert)$ and $\mathcal{O}(\vert \mathcal{C} \vert)$ time, respectively.

With the above transformations, the resulting QUBO problem is of the form
\begin{subequations} 
\allowdisplaybreaks
\begin{align}
    &\underset{\mathbf{x}}{\text{minimize}} \ \sum_{i \in \mathcal{L}} k_i^L x_i^L+ \sum_{j \in \mathcal{C}} k_j^C x_j^C  \label{design_qubo_obj}
    \\&+ M_1\left(\sum_{i \in \mathcal{L}}x_i^L -1 \right)^{\!\!2} \label{design_qubo_1}
    \\&+ M_2 \left(\sum_{j \in \mathcal{C}}x_j^C -1 \right)^{\!\!2} \label{design_qubo_2}
    \\&+ M_3 \sum_{i \in \mathcal{L}} \sum_{j \in \mathcal{C}}\left(3z_{ij}+x_i^L x_j^C - 2 x_i^L z_{ij} - 2 x_j^C z_{ij} \right) \label{design_qubo_3}
    \\&+ M_4 \left(\sum_{i \in \mathcal{L}} \sum_{j \in \mathcal{C}} z_{ij} - 1\right)^{\!\!2} \label{design_qubo_4}
    \\&+M_5 \left(\left( \frac{\kappa}{2 \pi f_{\textrm{sw}}} \right)^{\!2}  - \sum_{i \in \mathcal{L}} \sum_{j \in \mathcal{C}} z_{ij} L_i C_j \right) \label{design_qubo_5}
    \\&+ M_6 \left(\left( \frac{\kappa}{2 \pi f_{\textrm{sw}}} \right)^{\!2} - \sum_{i \in \mathcal{L}} \sum_{j \in \mathcal{C}} z_{ij} L_i C_j \right)^{\!\!2} \label{design_qubo_6} \,,
\end{align}%
\label{design_problem_QUBO}%
\end{subequations}%
where $M_{1,\ldots,6}$ are appropriately selected non-negative penalty weights. 

The number of qubits required to encode~\eqref{design_problem_QUBO} corresponds to the number of decision variables and is $N^Q=\vert \mathcal{L} \vert + \vert \mathcal{C} \vert + \vert \mathcal{L}\vert \times \vert \mathcal{C} \vert$. The number of logical gates required to implement QAOA depends on the number of layers $p$ and the number of linear and quadratic terms in the QUBO Hamiltonian of interest. For problem \eqref{design_problem_QUBO}, there are $N^L=N^Q$ linear and $N^{Qu}=\vert \mathcal{L} \vert\times\vert\mathcal{C}\vert+\vert \mathcal{L}\vert^2\times\vert\mathcal{C}\vert+\vert\mathcal{L}\vert\times\vert\mathcal{C}\vert^2$ quadratic terms. Implementing QAOA requires $N^Q$ Hadamard gates, $p\cdot N^Q$ $R_X$ gates for the mixer layers, as well as $p \cdot N^Q$ $R_Z$ and $p\cdot N^{Qu}$ $R_{ZZ}$ gates for the cost layers. Before execution on a QPU, the circuit must be transpiled to match the hardware’s native gate set and qubit connectivity. This process, together with the application of error suppression techniques, typically increases the overall gate count due to gate decomposition, routing overhead, and the additional operations required for noise suppression. Note that this is a hardware- and transpilation software-dependent process and it is therefore generally difficult to precisely characterize.

Finally, it is important to note that the optimal (or any feasible) solution of~\eqref{design_problem} will make terms~\eqref{design_qubo_1}--\eqref{design_qubo_4} exactly zero. However, the terms~\eqref{design_qubo_5} and~\eqref{design_qubo_6} will not necessarily be zero, as the unbalanced penalization is an approximate method for encoding inequality constraints. Therefore, the optimal solution of the original MINLP problem~\eqref{design_problem} will, in general, differ from that of the QUBO formulation~\eqref{design_problem_QUBO}. As discussed in Section~\ref{sec::Results}, careful selection of the penalty weights increases the likelihood that the optimal solution to the QUBO problem corresponds to a feasible and near-optimal (or optimal) solution of the original MINLP~\eqref{design_problem}.

\section{Numerical and Experimental Results}
\label{sec::Results}

\subsection{Experimental Setup}
\label{subsec:setup}

Quantum algorithms were implemented using PennyLane 0.40.0~\cite{pennylane} and simulations presented in Section~\ref{simulated_qaoa} were executed using an ideal statevector simulator on a MacBook Pro (Apple M3 Pro, 11-core CPU, 18 GB memory) running macOS Sequoia 15.5. The variational parameters of the QAOA circuit were optimized using the \emph{Adam} optimizer~\cite{adam_2017} with a step size of $10^{-3}$ and a limit of $2000$ iterations. Vectors $\boldsymbol  \beta$ and $\boldsymbol \gamma$ were initialized to $0.01$. Due to the small size of the problem instances, the sets of optimal, feasible, and infeasible solutions were identified by complete enumeration.

In addition to the simulated QAOA, we also conducted experiments on a real QPU. Specifically, we used \texttt{ibm\_kingston}, a 156-qubit Heron R2 quantum processor~\cite{Heron_QPU}. The native gate set of this QPU includes $CZ$, $I$, $R_X$, $R_Z$, $R_{ZZ}$, $SX$, and $\sigma^x$.\footnote{$CZ$ is the controlled-Z gate and $SX$ is the $\sqrt{\sigma^x}$ gate.} To enable efficient execution, all quantum circuits were transpiled and optimized using both Qiskit's native transpiler and Q-CTRL’s advanced circuit optimization tools~\cite{QCTRL}, which aim to minimize circuit depth and improve hardware performance. The corresponding results are presented in Section~\ref{ibm_qaoa}.

When a measurement operation is performed on all qubits, the decision variables of~\eqref{design_problem_QUBO} are associated with the qubits in the resulting computational basis state as $\ket{x_0^L,\!\dotsc\!,x_{\vert \mathcal{L} -1\vert}^L,x_0^C,\!\dotsc\!,x_{\vert \mathcal{C}-1 \vert}^C,z_{00},\!\dotsc\!,z_{\vert \mathcal{L}-1 \vert \vert \mathcal{C}-1 \vert}}$.

\begin{table}[!t]
  \footnotesize                                 
  \setlength{\tabcolsep}{2pt}                   
  \caption{Component values and costs used in this study.}
  \centering
  \begin{tabular*}{\linewidth}{@{\extracolsep{\fill}}cccc}
    \toprule
    \multicolumn{4}{c}{\textbf{Instance 1}} \\ \midrule
    \textbf{Inductance (µH)} & \textbf{Capacitance (µF)}
        & \textbf{Inductor cost (€)} & \textbf{Capacitor cost (€)} \\ \midrule
      10 &  54  & 0.5 & 1   \\
      22 & 115  & 0.9 & 1.5 \\ \midrule
    \multicolumn{4}{c}{\textbf{Instance 2}} \\ \midrule
    \textbf{Inductance (µH)} & \textbf{Capacitance (µF)}
        & \textbf{Inductor cost (€)} & \textbf{Capacitor cost (€)} \\ \midrule
      10 &  54  & 0.5 & 1   \\
      22 & 115  & 0.9 & 1.5 \\
      47 &      & 1.5 &     \\ \midrule
    \multicolumn{4}{c}{\textbf{Instance 3}} \\ \midrule
    \textbf{Inductance (µH)} & \textbf{Capacitance (µF)}
        & \textbf{Inductor cost (€)} & \textbf{Capacitor cost (€)} \\ \midrule
      10 &  54  & 0.5 & 1   \\
      22 & 115  & 0.9 & 1.5 \\
      47 & 253  & 1.5 & 2.5 \\ \bottomrule
  \end{tabular*}
  \label{tab:datasets}
\end{table}

\subsection{Example Instances}

We demonstrate the solution of the design problem with the QAOA algorithm using three instances that are presented in Table~\ref{tab:datasets}. The instances have the following optimal and feasible solutions: \looseness=-1
\begin{itemize}
    \item Instance 1: the optimal solution is $x_1^L=1$, $x_0^C=1$, and $z_{10}=1$, which translates to $\ket{01100010}$ or, for brevity, $\ket{98}$. The feasible solutions are $\ket{81}$ and $\ket{148}$.
    \item Instance 2: the optimal solution is $\ket{648}$ (corresponding to the same optimal inductor and capacitor values as in Instance 1) and the feasible solutions are $\ket{321}$, $\ket{386}$, $\ket{580}$, and $\ket{1104}$.
    \item Instance 3: the optimal solution is $\ket{10272}$ (\ie{} the same optimal values as in the other instances) and the feasible solutions are $\ket{4609}$, $\ket{5122}$, $\ket{6148}$, $\ket{8712}$, $\ket{9232}$, $\ket{16960}$, and $\ket{17536}$.
\end{itemize}

\begin{figure}[t]
  \centering
  \subfigure[Instance 1 (8 variables). $M_5=8.507$, $M_6=1.877$.]{\includegraphics[width=1\linewidth]{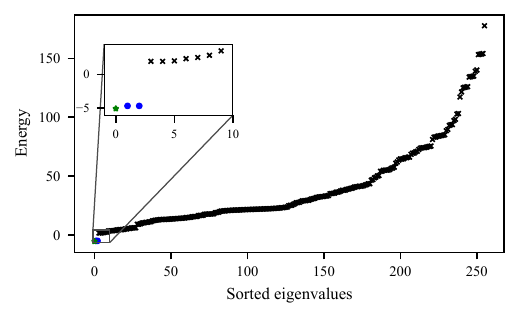}%
    \label{fig:eigen_dataset0}}
  \vfill
  \subfigure[Instance 2 (11 variables). $M_5=4.033$, $M_6=0.411$.]{\includegraphics[width=1\linewidth]{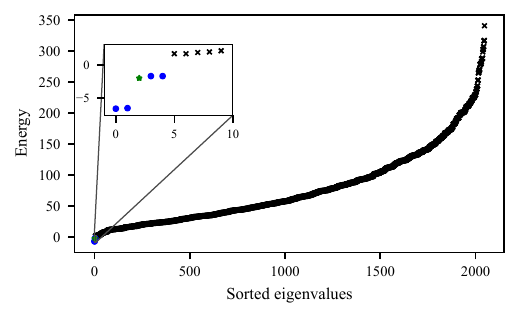}%
    \label{fig:eigen_dataset1}}
  \vfill
  \subfigure[Instance 3 (15 variables). $M_5=1.756$, $M_6=0.079$.]{\includegraphics[width=1\linewidth]{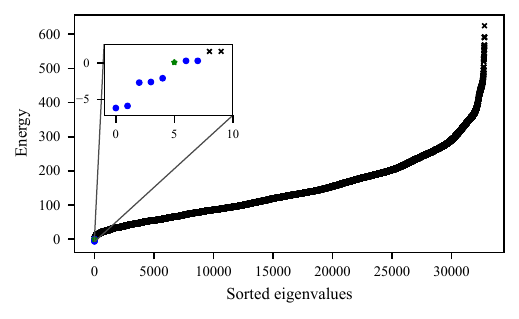}%
    \label{fig:eigen_dataset2}}
  \\[-0.5ex]               
  \caption{Eigenvalue distributions of the Ising formulation of problem~\eqref{design_problem_QUBO}. The inset shows the $10$ lowest-energy eigenvalues. The green star corresponds to the optimal solution of problem~\eqref{design_problem}, blue circles correspond to feasible solutions, and black crosses represent infeasible solutions.}
  \label{fig:eigen}
\end{figure}

\subsection{Weight Tuning}
\label{sec:tuning}

The quality of the QUBO problem solution depends on the identification of appropriate values for the penalty weights such that states corresponding to infeasible solutions are assigned higher energy levels of the cost Hamiltonian (Ising model resulting from~\eqref{design_problem_QUBO}) compared to states corresponding to feasible solutions and that, ideally, the lowest-energy state corresponds to the optimal solution. Given that the problem instances of Table \ref{tab:datasets} are small, penalties can be tuned through an exact procedure. We first set $M_1 = M_2 = M_3 = M_4 = 5$. Then, we formulate a mixed-integer linear program (MILP) that exhaustively accounts for the energy of each of the possible solutions. The objective function represents the maximization of the gap between the feasible solution with the highest energy and the infeasible solution with the minimum energy. Additionally, it is required that all feasible solutions correspond to a lower energy level compared to infeasible solutions.\looseness=-1 

The eigenvalue distributions for the three datasets are given in Fig.~\ref{fig:eigen}. For Dataset 1, the Ising Hamiltonian encodes the optimal solution as the one with the lowest energy. This is not the case for Datasets 2 and 3. However, all feasible solutions correspond to energy levels that are lower than those of infeasible solutions. These results are expected and are in line with the literature~\cite{Barrera_2024}, as the use of unbalanced penalization causes the QUBO problem~\eqref{design_problem_QUBO} to represent only an approximate encoding of the MINLP problem~\eqref{design_problem}.

\subsection{Simulated QAOA}
\label{simulated_qaoa}

\subsubsection{QAOA Landscape}

The quality of solutions obtained through the QAOA depends heavily on the number of layers $p$ and the successful identification of the parameter vectors $\boldsymbol \beta$ and $\boldsymbol \gamma$ for the specific problem instance. It should be emphasized that finding suitable QAOA parameters is itself a challenging problem~\cite{Barrera_2025}. A common approach involves using a classical optimizer to iteratively update these parameters in order to minimize the expected value of the cost Hamiltonian.

\begin{figure}[t]
    \centering
    \includegraphics[width=1\linewidth]{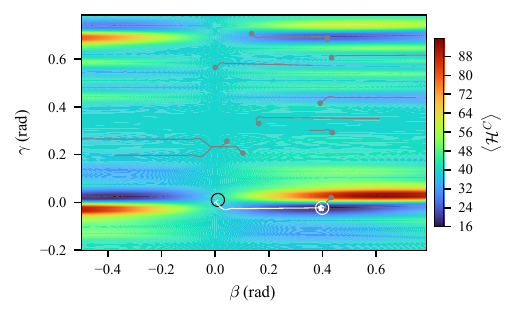}
    \caption{QAOA landscape of Instance 1 for $p=1$ using a $500\times 500$ grid for $\beta$ and $\gamma$ between $-\frac{\pi}{4}$ and $\frac{\pi}{4}$ (axes are truncated for legibility). The black and white circles indicate the start and end points of Adam optimizer for the initialization described in Section~\ref{subsec:setup} and the white line its trajectory. The white star denotes the point of minimum energy observed in the grid. Gray points indicate the start point of $10$ random initializations and gray lines the trajectory that was followed by the Adam optimizer.}
    \label{fig:qaoa_landscape}
\end{figure}

For $p=1$, it is possible to explicitly visualize the QAOA parameter landscape. We discretize both $\beta$ and $\gamma$ in the range $\left[-\frac{\pi}{4}, \frac{\pi}{4} \right]$ using $500$ equidistant grid points. The resulting QAOA landscape for Instance 1 is shown in Fig.~\ref{fig:qaoa_landscape}. In the numerical experiments, the gradient-based optimizer Adam is used, which is sensitive to the initial parameter values. In particular, as observed, random initializations of $\beta$ and $\gamma$ within the range $\left[0, \frac{\pi}{4} \right]$ often lead the optimizer to become trapped in local minima or plateaus. In contrast, initializing them near zero results in identifying parameters that correspond to the minimum expected energy over the landscape. The landscapes for Instances 2 and 3 exhibit similar characteristics, motivating our decision to initialize $(\boldsymbol{\beta}, \boldsymbol{\gamma})$ near zero in our numerical experiments. \looseness=-1

\subsubsection{Solution of Problem~\eqref{design_problem_QUBO} by Simulated QAOA} 
\begin{figure}[t]
    \centering
    \includegraphics[width=1\linewidth]{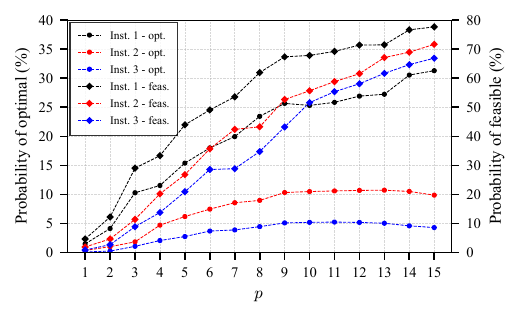}
    \caption{Probability of obtaining the true optimal or a feasible solution of problem \eqref{design_problem} by solving problem \eqref{design_problem_QUBO} in all three instances for different values of $p$.}
    \label{fig:probs_3dinstances}
\end{figure}

The cost Hamiltonian the ground state of which is sought corresponds to problem~\eqref{design_problem_QUBO}, however, we evaluate the obtained results against problem~\eqref{design_problem} since it is the problem of interest.
Figure~\ref{fig:probs_3dinstances} shows the probability of obtaining the true optimal solution or a feasible solution of problem~\eqref{design_problem} as the number of QAOA layers $p$ increases, after optimizing parameters $(\boldsymbol \beta, \boldsymbol \gamma)$ for Instances~1--3. As can be seen, the probability of obtaining a feasible solution increases monotonically with $p$ in all three instances. This is explained by the fact that deeper circuits are more expressive than shallow ones, allowing for further minimization of the expected value of the cost Hamiltonian. This, in turn, amplifies the probability of observing lower-energy solutions that correspond to feasible and optimal solutions, as discussed in Section~\ref{sec:tuning}. This observation is further corroborated by the results presented in Table~\ref{tab:solution-details}, which records the probability and status of the most probable solution as $p$ increases. Notably, for all three considered instances, the highest probability corresponds to a valid solution for $p \geq 3$.

However, the probability of obtaining the true optimal solution, \ie{} the solution of problem~\eqref{design_problem}, increases monotonically with $p$ only for Instance 1. For Instances 2 and 3, this probability initially increases with $p$, but then gradually decreases as the number of layers continues to grow. As discussed in Section~\ref{sec:tuning}, this behavior is due to the inexactness of the unbalanced penalization method used to encode~\eqref{design_nlc}. As a result, the ground state of problem~\eqref{design_problem_QUBO}, which QAOA successfully amplifies, does not correspond to the optimal solution of the original problem~\eqref{design_problem}. Thus, the improved performance of QAOA with  increasing $p$ leads to the amplification of a suboptimal but feasible solution of problem \eqref{design_problem}, \ie{} $\ket{580}$ ($\ket{5122}$) is promoted instead of $\ket{648}$ ($\ket{10272}$) in Instance 2 (Instance 3). Nonetheless, as explained in Section~\ref{sec:merit}, the true optimal solution (\ie{} the solution of problem~\eqref{design_problem}) can be recovered with substantial probability. \looseness=-1

These results highlight the importance of problem encoding. While deeper and properly trained QAOA ansatzes are theoretically expected to increase success probability, in practice, when the constrained problem is not equivalent to its Ising representation, adding layers will improve convergence to the QUBO ground state but not necessarily to the true optimum of the problem of interest.\looseness=-1

\begin{table*}[]
\centering
\footnotesize                                 
\setlength{\tabcolsep}{2pt}    
\caption{Most probable solution characteristics}
\label{tab:solution-details}
\begin{tabular*}{\linewidth}{@{\extracolsep{\fill}}llccccccccccccccc}
\toprule
$p$ &  & 1 & 2 & 3 & 4 & 5 & 6 & 7 & 8 & 9 & 10 & 11 & 12 & 13 & 14 & 15 \\ \midrule
\multirow{3}{*}{\rotatebox{90}{\textbf{Dataset 1}}} & \textbf{Solution} & $\ket{81}$ & $\ket{98}$ & $\ket{98}$ & $\ket{98}$ & $\ket{98}$ & $\ket{98}$ & $\ket{98}$ & $\ket{98}$ & $\ket{98}$ & $\ket{98}$ & $\ket{98}$ & $\ket{98}$ & $\ket{98}$ & $\ket{98}$ & $\ket{98}$ \\ \cmidrule(l){2-17} 
 & \textbf{Solution status} & feas. & opt. & opt. & opt. & opt. & opt. & opt. & opt. & opt. & opt. & opt. & opt. & opt. & opt. & opt. \\ \cmidrule(l){2-17} 
 & \textbf{Prob. (\%)} & 1.68 & 4.14 & 10.29 & 11.55 & 15.39 & 18.01 & 19.95 & 23.44 & 25.66 & 25.32 & 25.85 & 26.92 & 27.24 & 30.54 & 31.29 \\ \midrule
\multirow{3}{*}{\rotatebox{90}{\textbf{Dataset 2}}} & \textbf{Solution} & $\ket{320}$ & $\ket{320}$ & $\ket{386}$ & $\ket{1104}$ & $\ket{1104}$ & $\ket{386}$ & $\ket{580}$ & $\ket{580}$ & $\ket{580}$ & $\ket{580}$ & $\ket{580}$ & $\ket{580}$ & $\ket{580}$ & $\ket{580}$ & $\ket{580}$ \\ \cmidrule(l){2-17} 
 & \textbf{Solution status} & infeas. & infeas. & feas. & feas. & feas. & feas. & feas.* & feas.* & feas.* & feas.* & feas.* & feas.* & feas.* & feas.* & feas.* \\ \cmidrule(l){2-17} 
 & \textbf{Prob. (\%)} & 0.43 & 1.11 & 2.56 & 4.79 & 6.60 & 8.89 & 10.85 & 10.94 & 14.30 & 15.69 & 17.05 & 18.28 & 21.27 & 22.56 & 24.77 \\ \midrule
\multirow{3}{*}{\rotatebox{90}{\textbf{Dataset 3}}} & \textbf{Solution} & $\ket{12800}$ & $\ket{4609}$ & $\ket{6148}$ & $\ket{6148}$ & $\ket{6148}$ & $\ket{6148}$ & $\ket{6148}$ & $\ket{6148}$ & $\ket{5122}$ & $\ket{5122}$ & $\ket{5122}$ & $\ket{5122}$ & $\ket{5122}$ & $\ket{5122}$ & $\ket{5122}$ \\ \cmidrule(l){2-17} 
 & \textbf{Solution status} & infeas. & feas. & feas. & feas. & feas. & feas. & feas. & feas. & feas.* & feas.* & feas.* & feas.* & feas.* & feas.* & feas.* \\ \cmidrule(l){2-17} 
 & \textbf{Prob. (\%)} & 0.24 & 0.46 & 1.25 & 2.13 & 3.18 & 4.39 & 4.52 & 5.40 & 6.86 & 9.12 & 10.14 & 11.05 & 12.31 & 14.00 & 15.20 \\ \bottomrule
\end{tabular*}
\begin{flushright}
  \footnotesize “opt.” = true optimal of problem \eqref{design_problem}, “infeas.” = infeasible, “feas.” = feasible, “feas.*” = optimal of problem \eqref{design_problem_QUBO}, encoding-based minimum-energy solution.
\end{flushright}
\end{table*}

\begin{figure}[t]
  \centering
  \subfigure[CoP]{\includegraphics[width=1\linewidth]{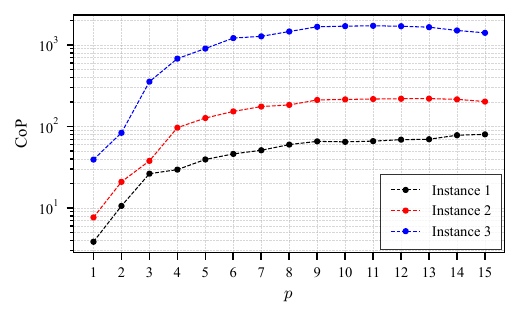}%
    \label{fig:cop}}
  \vfill
  \subfigure[TTS]{\includegraphics[width=1\linewidth]{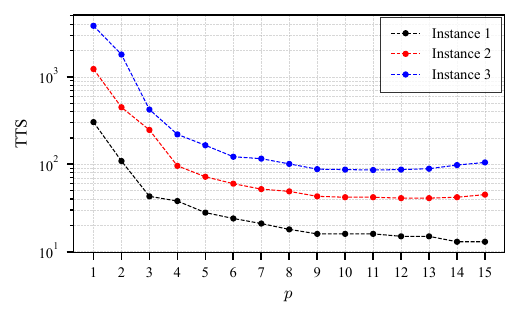}%
    \label{fig:tts}}
  \\[-0.5ex]               
  \caption{Figures of merit in all three instances for different values of $p$.}
  \label{fig:merit}
\end{figure}

\subsubsection{Figures of Merit} 
\label{sec:merit}

Although benchmarking QAOA with classical optimization techniques for the prototypical design problem falls outside of the scope of this paper, the calculation of different figures of merit can still provide valuable insight into performance characteristics and limitations. Furthermore, it is worth pointing out that the establishment of fair benchmarking practices is a challenging and active area of research~\cite{Bucher_2024, koch_2025}.

First, similar to classical metaheuristics, QAOA may be a useful solver only if the probability of obtaining the optimal solution of the problem is substantial. Therefore, it is pertinent to compare the probability that QAOA discovers the optimal solution with random guessing over the set of possible solutions. This is quantified by the coefficient of performance (CoP) defined as 
\begin{equation}
    \textrm{CoP}= \frac{\pi^*}{\pi^r} \,,
\end{equation}
where $\pi^*$ is the probability that QAOA returns the true optimal solution and $\pi^r=\frac{1}{2^n}$ is the probability of randomly guessing it, with $n$ being the number of qubits. 

Figure~\ref{fig:cop} shows the CoP for the three instances as $p$ increases. It can be observed that, for any number of QAOA layers $p$, the CoP increases with increasing problem size. This is explained by the fact that as the number of problem variables increases, the probability of obtaining the optimal bitstring by random guessing diminishes exponentially. We also note that although CoP for Instance 1 increases monotonically with increasing $p$, this is not the case for Instances 2 and 3, as a result of the decreasing probability to obtain the true optimum of problem \eqref{design_problem} for these two cases. 

\begin{figure*}[t!]
    \centering
    \includegraphics[width=1\linewidth]{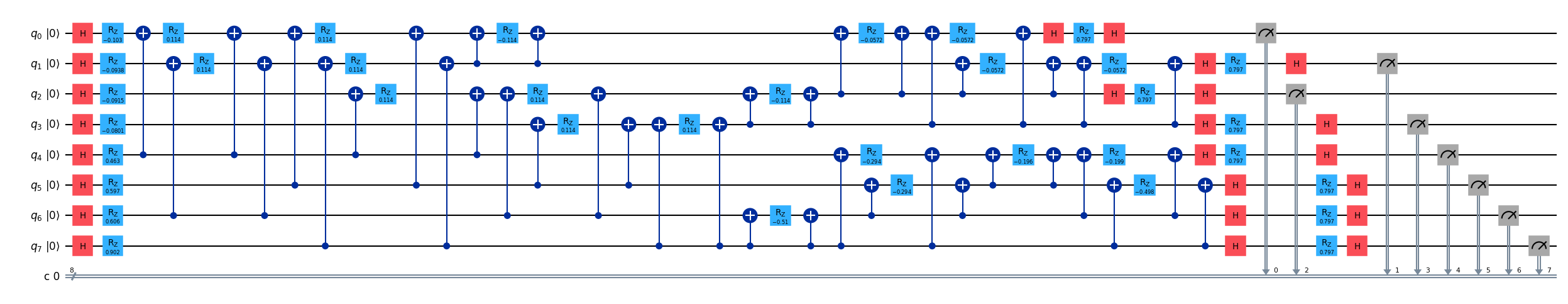}
    \caption{Decomposed QAOA quantum circuit for Instance 1 and $p=1$ before transpilation.}
    \label{fig:circuit_p1}
\end{figure*}

Another important figure of merit of any solution algorithm is the run-time required to obtain the optimal solution with a certain probability. Unlike statevector simulators, which calculate the probability of measuring a certain state exactly, real quantum hardware relies on performing a large number of measurements (shots) to obtain a representative set of samples. A commonly used proxy for the computational time required to solve the optimization problem using a probabilistic algorithm is the time-to-solution (TTS), defined as
\begin{equation}
    \textrm{TTS} = \Lambda \ceil*{\frac{\ln(1-\alpha)}{\ln(1-\pi^*)}} \,,
\end{equation}
where $\pi^*$ is the probability that QAOA returns the optimal solution, $\alpha$ is the probability threshold, and $\Lambda$ is the time to obtain a single sample. Alternatively, $\Lambda$ can be estimated as the number of circuit layer operations~\cite{bucher_2025}, or the sum of the variational parameters~\cite{Zhou_2020}. In this study, we assume that $\Lambda = 1$ and $\alpha=0.99$. Hence, TTS stands for the number of shots required to observe the optimal solution at least once with a probability of $99\%$. Note that the number of shots reported in this paper should be interpreted as a theoretical lower bound when experimenting with real hardware, given that our simulations do not account for noise.

The TTS for the three instances for increasing values of $p$ is shown in Fig.~\ref{fig:tts}. As expected, TTS decreases monotonically for Instance 1, whereas this trend is not observed for Instances 2 and 3. However, for these two instances, a significant improvement in TTS can be observed when $p \geq 4$. These theoretical findings indicate that the true optimal solution for all three problem instances may be observed at least once with a reasonable number of shots on a real QPU, even when employing shallow QAOA circuits.




\subsection{Solution of Problem \eqref{design_problem_QUBO} on an IBM Quantum Computer}
\label{ibm_qaoa}

We implemented Instance 1 on \texttt{ibm\_kingston}. For $p \in \left[ 1, 10 \right]$, we use the optimized circuit parameter vectors $\boldsymbol \beta$ and $\boldsymbol \gamma$ that were found by training the QAOA ansatz using the statevector simulator. For each experiment, $4000$ shots were used, which exceeds the number of shots estimated in Fig.~\ref{fig:tts} to observe the true optimal at least once with a probability $99\%$ by at least a factor of $13$.

\begin{figure}[t]
    \centering
    \includegraphics[width=1\linewidth]{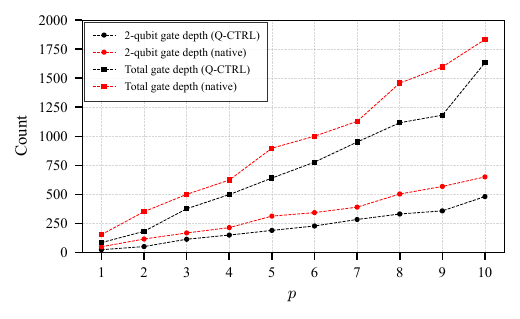}
    \caption{Two-qubit gate count and total circuit depth after transpilation of the QAOA ansatz corresponding to Instance 1 using the \enquote{native} Qiskit transpiler and the advanced Q-CTRL transpiler for different values of $p$.}
    \label{fig:gate_count}
\end{figure}

\begin{figure}[t]
    \centering
    \includegraphics[width=1\linewidth]{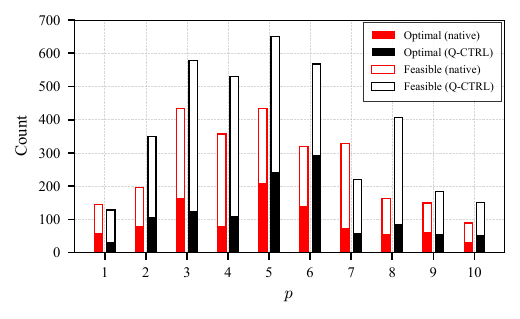}
    \caption{Comparison of Instance 1 optimal and feasible solution counts from QAOA executions on the \texttt{ibm\_kingston} QPU with 4000 shots, using both the \enquote{native} Qiskit transpiler (with no error mitigation/supression) and the advanced Q-CTRL transpiler (with error mitigation/supression) for different values of $p$.}
    \label{fig:counts_error_suppression}
\end{figure}

For $p=1$, the trained quantum circuit is shown in \mbox{Fig.~\ref{fig:circuit_p1}}. Comparing the gate set that is used in Fig. \ref{fig:circuit_p1} with the gate set in a circuit that implements QAOA at a high level, such as in \mbox{Fig. \ref{fig:didactic_example_circuit}} in Appendix \ref{app:qaoa-example}, the $R_{ZZ}$ gates were further decomposed in terms of $R_Z$ and $CX$ gates through the identity $R_{ZZ}(\theta)=CX (I \otimes R_Z(\theta))CX$, and the $R_X$ gate was decomposed in terms of $R_Z$ and $H$ gates using the identity $R_{X}(\theta)=H R_{z}(\theta) H$. Before such a quantum circuit can be executed on a QPU, it must be transpiled to use the native gate set and respect the physical topology of the given architecture. A key metric that characterizes the reliable execution of a quantum algorithm on a QPU is the circuit depth, defined as the number of sequential layers of quantum gates required to perform the computation. In general, deeper circuits are more error-prone, as the errors associated with each quantum gate accumulate over a chain of operations due to coherence times of the qubits. Two-qubit gates have higher implementation error rates than single-qubit gates, which makes it important to track their count separately. Figure~\ref{fig:gate_count} shows the results for the circuits transpiled using two different approaches, namely the native Qiskit transpiler and the Q-CTRL transpiler. As mentioned in Section~\ref{subsec:setup}, the Qiskit transpiler optimizes quantum circuits for hardware execution, while Q-CTRL tailors those circuits to be more robust against noise and errors. As expected, the depth of the transpiled circuit increases with the number of QAOA layers $p$, in a practically linear manner. Interestingly, Q-CTRL is capable of reducing both the circuit depth and the number of 2-qubit gates, which highlights the importance of effective quantum circuit pre-processing for execution on real quantum hardware. \looseness=-1

Figure \ref{fig:counts_error_suppression} compares the number of samples corresponding either to the optimal or to a feasible bitstring across different QAOA circuit depths $p$. Across all tested depths, the use of Q-CTRL increased the average count of optimal bitstrings by 22.95\% and the count of feasible solutions by 48.75\% relative to results obtained using Qiskit’s standard transpilation. Notably, the highest probability of obtaining the optimal solution for Instance 1 was observed at $p=6$ when using Q-CTRL, demonstrating that effective error mitigation and suppression techniques can enhance the quality and reliability of quantum optimization outcomes in practice.

\begin{figure}[t]
    \centering
    \includegraphics[width=1\linewidth]{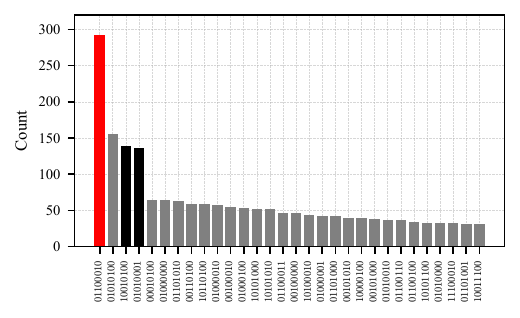}
    \caption{Frequency of the top 30 bitstrings observed over $4000$ shots for Instance~1 executed on the \texttt{ibm\_kingston} QPU with depth $p = 6$. The red bar highlights the optimal bitstring, while the black bars correspond to feasible solutions.
}
    \label{fig:rank_p6}
\end{figure}

Analyzing the results shown in Fig. \ref{fig:counts_error_suppression} using Q-CTRL in more detail, Table~\ref{tab:counts_ibm} presents the rank of optimal and feasible bitstrings across different QAOA depths $p$. In all cases, the probability of obtaining the true optimal bitstring is non-negligible, whereas the probability of obtaining a feasible bitstring is considerably higher. The fact that the hardware implementation of QAOA consistently amplifies the optimal solution is indicated by the CoP that ranges from $2.11$ for $p=1$ to $18.75$ for $p=6$. Notably, the performance of QAOA on quantum hardware does not exhibit the monotonic improvement with increasing depth seen in simulation, which can be attributed to hardware noise and the fact that the ansatz parameters were trained using an ideal statevector simulator. Interestingly, for $p=6$ and $p=10$, the true optimal bitstring is ranked first among the $256$ possible outcomes. Moreover, in all cases except for $p=1$ and $p=7$, the most frequently measured bitstring corresponds either to the true optimal or a feasible solution of problem~\eqref{design_problem}. As an example, Fig.~\ref{fig:rank_p6} shows the counts of the top-30 bitstrings for $p=6$. Given that it suffices that the true optimal appears at least once in the set of samples, these results suggest that QAOA can reliably generate both the true optimal and feasible solutions of problem~\eqref{design_problem} also when executed on a real QPU. 

Hence, the power electronics design problem presented in Section~\ref{sec::DesPE} has been successfully implemented and solved on a QPU, marking a noteworthy---if not symbolic---milestone for the power electronics community. Building on the insights gained from this exercise, we share our perspective and outlook on the future role of quantum computing in power electronics. \looseness=-1

\begin{table}[t]
\centering
\footnotesize
\setlength{\tabcolsep}{2pt}        
\caption{Counts of the optimal and feasible bitstrings across circuit depths $p$ from execution on the \texttt{ibm\_kingston} QPU with 4000 shots and Q-CTRL. The rank of the optimal (feasible) bitstring corresponds to its position among all $2^8$ bitstrings when sorted by observed frequency in descending order, with rank 1 being the most frequently measured bitstring.}
\label{tab:counts_ibm}
\begin{tabular}{@{}ccccc@{}}
\toprule
$p$ & \shortstack{\textbf{Optimal} \\ \textbf{bitstring count}}  & \shortstack{\textbf{Feasible} \\ \textbf{bitstring count}} & \shortstack{\textbf{Rank of} \\ \textbf{optimal bitstring}} & \shortstack{\textbf{Rank of} \\ \textbf{feasible bitstrings}} \\ \midrule
1 & 33 & 128 & 29 & 9, 16\\
2 & 106 & 349 & 2 & 1, 8\\
3 & 126 & 579 & 4 & 1, 3\\
4 & 111 & 531 & 4 & 1, 2\\
5 & 244 & 650 & 2 & 1, 3\\
6 & 293 & 568 & 1 & 3, 4\\
7 & 60  & 220 & 9 & 2, 20\\
8 & 85  & 406 & 6 & 1, 5\\
9 & 57  & 183 & 4 & 1, 73\\
10 & 52 & 151 & 1 & 2, 3\\ \bottomrule
\end{tabular}
\end{table}

\section{Vision}
\label{sec::Vision}



\subsection{Current Limitations and Anticipated Advancements in Quantum Computing}
\label{sec::ChLim}

The (simplified) design problem introduced in Section~\ref{sec::QuantPE} offers a glimpse into the potential role of quantum computing in power electronics. Quantum computing, particularly in conjunction with classical embedded algorithms, could play a transformative role in this field, as it is well suited to addressing complex multiobjective optimization problems with discrete and nonlinear constraints, \ie{} features that are characteristic of many design tasks in power electronics. However, as evidenced from the relative simplicity of the chosen power electronics example, the application of quantum computing to problems that feature a complexity of practical relevance remains a significant challenge and is still some way off.

The reason for this is two-fold. Specifically, to translate the transformative potential of quantum computing into tangible impact for power electronics applications, two key conditions must be met. First, one needs to be able to cast different power electronics design problems as combinatorial optimization problems, and, subsequently, reformulate them such that they are amenable to solution by a quantum optimization algorithm. This implies either devising a QUBO formulation or using embeddings that do not rely on the QUBO paradigm and may allow for efficiently capturing higher order interactions, such as the parity architecture \cite{Dominguez_2023, parity_original, parity_2}. The power electronics community has the ability to satisfy this requirement after acquiring a considerable degree of familiarity with the rather demanding mathematical background. Second, access to powerful quantum computing hardware capable of dealing with problems of meaningful size is essential to demonstrate quantum advantage for a problem of industrial significance. This represents a bigger challenge because quantum computing technology is still in its infancy. However, as discussed in Section \ref{sec:quantum_costs}, NISQ quantum hardware is essentially readily accessible to enable exploratory research in the power electronics domain, whereas the quantum computing community is moving forward in addressing these limitations, and advancements are anticipated, as outlined below.


Over the next decade, quantum computing will transition from NISQ devices to early fault-tolerant quantum computers (FTQC) which will be capable of reliably executing deeper, more complex quantum circuits due to the inclusion of quantum error correction codes\footnote{Quantum error correction refers to a set of methods that redundantly encode quantum information to protect it against decoherence and other types of noise~\cite{Roffe_2019}.}. According to IBM’s updated hardware roadmap, FTQCs such as the IBM Starling (2029) and IBM Blue Jay (2033) are planned, aiming to support so-called error-corrected ``logical qubits” and deep quantum circuits for real-world applications~\cite{IBM_roadmap}. Other universal gate-based QC vendors, such as IonQ~\cite{IONQ_roadmap}, IQM~\cite{IQM_roadmap}, and Oxford Quantum Circuits~\cite{OQC_roadmap} have also published roadmaps targeting fault-tolerant QPUs through scalable qubit architectures within the decade. \looseness=-1

With the advent of these FTQC systems, hybrid variational algorithms such as the QAOA---which currently struggle on NISQ hardware due to circuit depth limitations---may demonstrate a practical quantum advantage for certain classes of hard combinatorial optimization problems. There is already growing evidence suggesting that QAOA can yield quantum speedups over classical approaches, with the extent of the advantage strongly depending on the problem structure. This potential ranges from low-order polynomial to even exponential speedup in special instances~\cite{basso2021quantum, boulebnane2024solving, montanaro2024quantum}. As the quantum computing ecosystem continues to mature and FTQC inches closer, there has been a push for the development of novel quantum algorithms in order to bridge the gap between current NISQ-friendly approaches and new ones capable of exploiting a modest number (\eg{} $100$--$200$) of error-corrected logical qubits. \looseness=-1

Realizing practical quantum advantage will ultimately depend on the evolution of the entire quantum computing stack. Solving problems of practical interest by harnessing a quantum speedup relies on realizing scalable and fault-tolerant architectures, alongside continued algorithmic developments to harness the computational capabilities of upcoming FTQC devices. Equally important is fostering close collaboration between quantum scientists, engineers, and domain experts in industry, to ensure that algorithmic and hardware advancements are aligned with real-world problem requirements, starting from prototypical applications today. Constructing a virtuous cycle between theoretical and technological developments and industrial applications is therefore pivotal. \looseness=-1

\subsection{Future of Quantum Computing in Power Electronics}
\label{sec::QuantumPE}

Given this anticipated progress in quantum computing, one can envision a future in which it supplants the exhaustive search methods that currently serve as the default approach for power electronics designers. Quantum computing could significantly reduce this computational burden by efficiently exploring large solution spaces. Moreover, the effective generation of Pareto fronts enabled by quantum methods could facilitate the selection of optimal converter topologies or configurations---based on criteria such as efficiency, cost, complexity, and electromagnetic interference (EMI)---for a given application. Since some design variables, such as as the number of available switching states, are discrete by nature, the ability of quantum computing to handle combinatorial spaces becomes especially advantageous.

Extending this concept, quantum computing could also play a complementary role in simulation workflows. While it is not expected to directly accelerate classical simulation tools, such as PLECS or ANSYS, it may guide optimization tasks built around repeated simulations. For example, a high-dimensional design space of a converter could be cast as a combinatorial optimization problem, with a quantum optimizer identifying promising candidate configurations. These candidates would then be evaluated using classical simulations, reducing the total number of runs while retaining accuracy. It should be noted, however, that this hybrid quantum--classical approach depends on the development of large-scale, fault-tolerant quantum hardware and the availability of well-structured problem formulations. Nevertheless, it represents a compelling long-term pathway for integrating quantum computing into design, optimization, and simulation workflows in power electronics.

\begin{figure}[t!]
    \setlength{\figHeight}{3.cm}
    \setlength{\figWidth}{0.48\textwidth}
    \centering
    \includegraphics[width=\figWidth]{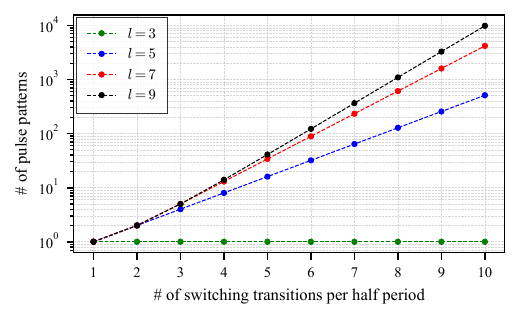}
    \caption{Number of possible $\ell$-level pulse patterns as a function of the number of switching transitions per half period (adapted from~\cite{KKG24c}).}
    \label{fig::noPatterns}
    \vspace{\figSpace}
\end{figure}

\begin{figure}[t!]
    \setlength{\figHeight}{3.cm}
    \setlength{\figWidth}{0.48\textwidth}
    \centering
    \includegraphics[width=\figWidth]{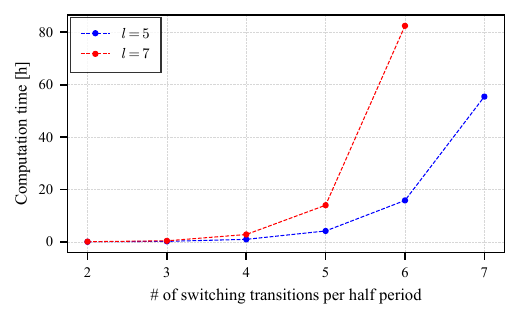}
    \caption{Computation times for $\ell$-level pulse patterns as a function of the number of switching transitions per half period (adapted from~\cite{KKG24c}).}
    \label{fig::compTime}
    \vspace{\figSpace}
\end{figure}

Beyond design problems, quantum computing may also be employed for offline optimal control and modulation tasks in power electronics, many of which are formulated as mixed-integer non-linear optimization problems. With regards to the latter, a representative example is the computation of pulse patterns with desirable characteristics, \eg{} patterns that produce the theoretically minimum load current harmonic distortions when applied to a power converter. The structure of these patterns depends on the number of switching transitions and voltage levels of the converter, leading to an exponential increase in the number of candidate solutions (see Fig.~\ref{fig::noPatterns}). Solving for the optimal pulse pattern thus requires addressing a complex mixed-integer, non-convex optimization problem, which is intrinsically difficult to solve. The alternative of exhaustively assessing each candidate pulse pattern quickly becomes infeasible for converters with a high number of voltage levels (see Fig.~\ref{fig::compTime}), thereby limiting the practical use of multilevel optimized pulse patterns and the benefits they could offer. Although some mathematical techniques can transform such problems such that their mixed-integer nature is masked, these methods do not guarantee fast convergence and often require repeated solving to obtain satisfactory results~\cite{LH17, PCK^+19}. \looseness=-1

Quantum computing, in contrast, could natively handle the discrete nature of such problems without requiring mathematical transformations that expand the feasible space and increase the complexity of the underlying optimization problems. This capability is particularly appealing for optimizing pulse patterns in multilevel converters, where conventional methods struggle to scale. Leveraging this capability would enable the practical use of multilevel optimized pulse patterns, which could significantly improve the load- and converter-friendly operation of power electronic systems by, \eg{} reducing harmonic distortions, increasing efficiency and output power, and extending system lifetime~\cite{GKK24, KKG24, DG24}.

A similar opportunity exists in control-related applications that involve solving large mixed-integer optimization problems offline. For example, in explicit model predictive control (MPC), the control policy is precomputed by solving a parametric optimization problem over the whole state space. This offline computation step can become computationally challenging as the problem size increases, especially when systems with integer variables or hybrid dynamics are considered, such as power electronic systems~\cite{GPM05, GPFM08, GPM08}. Another example is the generation of optimal switching trajectories for direct control strategies such as finite control set MPC (FCS-MPC), which can be designed to enforce desirable properties such as limit-cycle stability~\cite{XDL22}. In both scenarios, quantum computing---especially through QUBO-based formulations---offers the potential to accelerate these offline computationally intensive steps and expand the range of tractable problem sizes. This, in turn, would facilitate the adoption of optimal control techniques---known for their superior performance compared to conventional methods---for a wider range of power electronic applications.

Considering the above, it is straightforward to observe, however, a pattern where the anticipated benefits of applying quantum computing in power electronics are largely confined to offline computations performed during the design process, either at the converter level or within the control loop. The day when this technology will be used for online, real-time computations in power electronics seems to be very far ahead indeed. \looseness=-1

Beyond design as well as offline control and modulation applications, quantum computing is expected to create transformative synergies with emerging technologies. Quantum machine learning could enable predictive maintenance, anomaly detection, and control-policy discovery, while quantum-enhanced digital twins may accelerate parameter calibration, system diagnostics, and design optimization. Post-quantum cryptography could further secure communications in converter-dominated energy systems. Although these opportunities depend on the development of large-scale, fault-tolerant hardware and well-structured problem formulations, they illustrate the long-term potential of quantum computing to enhance both engineering workflows and the broader power electronics ecosystem.



\begin{figure*}[t]
    \centering
    \includegraphics[width=0.68 \linewidth]{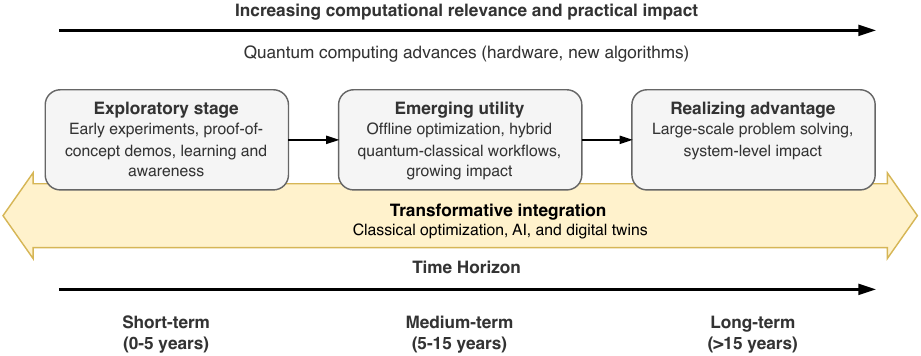}
    \caption{High-level conceptual roadmap illustrating the evolving role of quantum computing in power electronics.}
    \label{fig:roadmap_new}
\end{figure*}

\begin{table*}[t]
\centering
\caption{Short-, mid-, and long-term prospects for quantum computing in power electronics.}
\label{tab::roadmap}
\renewcommand{\arraystretch}{1.3}
\begin{tabular}{p{2.5cm} p{5.5cm} p{8.5cm}}
\toprule
{\bf Time horizon} & {\bf Expected quantum capabilities} & {\bf Implications for power electronics} 
\\
\midrule
{\bf Short-term} \newline ($0$--$5$ years) 
& NISQ devices with $\lesssim 10^3$ qubits, noisy results, accessed via cloud, limited error correction, restricted to problems of non-meaningful size for practical engineering use 
& \begin{itemize}[leftmargin=*]\vspace{-8pt}
    \item Proof-of-concept studies on simplified converter design and control problems
    \item Benchmarking workflows against classical approaches
    \item Educational and awareness-building use cases for power electronics researchers (academia and industry)
    \item Exploration of QUBO formulations of design/control problems
\end{itemize}
\\
\midrule
{\bf Mid-term} \newline ($5$--$15$ years) 
& Intermediate-scale devices with partial error correction, improved coherence times, early quantum advantage in structured optimization, and hybrid quantum--classical workflows 
& \begin{itemize}[leftmargin=*]\vspace{-8pt}
    \item Design-space exploration and Pareto-front generation for converters
    \item Offline optimization in control/modulation (\eg{} pulse pattern generation, explicit MPC)
    \item Quantum-assisted parameter sweeps in converter modeling and control
\end{itemize}
\\
\midrule
{\bf Long-term} \newline ($> 15$ years) 
& Large-scale (millions of qubits), fault-tolerant quantum computers capable of running complex algorithms (linear solvers, differential-equation solvers, quantum machine learning), error correction 
& \begin{itemize}[leftmargin=*]\vspace{-8pt}
    \item Integration into simulation workflows for accelerated optimization around nonlinear solvers
    \item Scalable multi-objective optimization for converter-dominated grids and system-level energy management
    \item Native handling of high-dimensional mixed-integer problems in real-world converters
\end{itemize}
\\
\midrule
\textbf{Cross-cutting opportunities} (across different time horizons) 
& Synergy with AI, digital twins, energy systems, and cybersecurity
& \begin{itemize}[leftmargin=*]\vspace{-8pt}
    \item Quantum machine learning for predictive maintenance, anomaly detection, or control policy discovery
    \item Quantum-enhanced digital twins for parameter calibration, diagnostics, and design optimization
    \item System-level optimization for converter-based grids
    \item Preparation for quantum-era security (\eg{} post-quantum cryptography for secure converter communications)
\end{itemize}
\\
\bottomrule
\end{tabular}
\end{table*}

\subsection{Short-, Mid-, and Long-Term Outlook}
\label{sec::OutlookPE}

Given the previous points, the trajectory of quantum computing in power electronics can be viewed over short-, mid-, and long-term horizons, as conceptually depicted in Fig.~\ref{fig:roadmap_new} and summarized in Table~\ref{tab::roadmap} in more detail. This combined outlook highlights not only the expected evolution of quantum hardware but also the potential implications for design, control, and simulation workflows in power electronics.

In the short term ($0$--$5$ years), quantum computing is primarily exploratory. Researchers can experiment with NISQ devices for educational purposes, small-scale proof-of-concept studies, and benchmarking of simplified converter design or offline control and modulation tasks. Although the problem sizes accessible today are limited and not yet practically meaningful for engineering applications, these early experiments are essential for building familiarity with quantum problem formulations, hybrid workflows, and the methodological foundations needed for future advances.

The mid term ($5$--$15$ years) could bring tangible computational advantages for specific classes of combinatorial optimization problems. As hardware improves and hybrid quantum--classical solvers mature, power electronics design-space exploration, Pareto-front generation for converter topologies, and offline optimization in control and modulation (\eg{} generation of multilevel pulse patterns or explicit MPC) may become tractable at previously prohibitive scales. Quantum-assisted parameter sweeps and optimization loops could significantly reduce computational bottlenecks, enabling more comprehensive evaluation of design alternatives and more informed engineering decisions.

Looking further ahead ($> 15$ years), the advent of large-scale, fault-tolerant quantum computers could enable the solution of high-dimensional mixed-integer problems at scales unattainable today. Beyond design and offline control, quantum computing may integrate with simulation workflows for accelerated optimization around nonlinear solvers and support system-level energy management in converter-based grids. Although real-time control remains a distant goal, the potential for scalable, practically relevant computations is significant.

Throughout these horizons, cross-cutting opportunities are expected to emerge at the intersection of quantum computing, artificial intelligence (AI), digital twins, and cybersecurity. AI has already become an increasingly important tool in, \eg{} the design and control of power electronic systems as well as in predictive maintenance and condition monitoring~\cite{ZBW21, LZLB22, LLL^+25, CCB^+25}. A potential synergy with quantum computing could amplify these capabilities by accelerating learning and optimization tasks, enabling AI models to explore larger design spaces and adapt to more complex converter behaviors. Quantum machine learning~\cite{Biamonte_2017, Acampora_2026} could enhance predictive maintenance or anomaly detection, while quantum-enhanced digital twins may accelerate parameter calibration, diagnostics, and design optimization. Additionally, post-quantum cryptography could secure communications in increasingly converter-based systems. Collectively, these developments illustrate a broad and transformative vision for the role of quantum computing in power electronics.

\section{Conclusions}
\label{sec::Concl}


This paper has explored the potential of quantum computing in power electronics, with particular emphasis on its ability to address computationally demanding mixed-integer optimization problems arising in the field. Despite being in a nascent stage, characterized by limited qubit counts, noise, and hardware constraints, quantum computing shows clear promise in tackling, among others, offline design and control problems in power electronics. Through a simplified case study involving the filter design of a dc–dc boost converter, we have demonstrated, for the first time, how such problems can be reformulated as QUBO problems and executed on quantum hardware, marking a symbolic yet meaningful milestone for the field. Although the chosen example is intentionally elementary, it serves as a key conceptual bridge between traditional optimization formulations and their quantum equivalents, and as a motivating case for early exploration within the power electronics community.

More broadly, this paper encourages the community to look beyond current technological limitations and engage with the emerging quantum paradigm. The insights gained here should not be seen as limitations but as a segue towards the computational avenues that quantum computing could, in time, open for power electronics. As hardware scales and fault tolerance improves, quantum methods may enable tractable solutions to complex, high-dimensional, multiobjective optimization problems that today remain beyond reach.

Ultimately, the picture envisioned in this paper is one in which quantum computing evolves from a conceptual curiosity into a transformative enabler of innovation in power electronics. Its impact will unfold gradually, first through exploratory studies and hybrid experimentation, and later through deeper integration into converter design, control, and beyond. This progression could unlock new opportunities for performance, efficiency, and scalability. Thus, early involvement of the power electronics community is crucial for shaping this transition and realizing the promise of quantum-enabled engineering.

\appendices

\section{Tutorial on Gate-based Quantum Computing}
\label{app:quantum-tutorial}

In this Appendix, we provide a brief overview of fundamental gate-based quantum computing notions. For an in-depth and rigorous treatment of these topics, the interested readers are referred to~\cite{quantum_book_1} and~\cite{quantum_book_2}.

\subsubsection{Qubits}

The fundamental unit of information in quantum informatics is the quantum bit or qubit. Whereas a classical bit can be either in state $0$ or $1$, the state $\ket{\psi}$ of a qubit is a superposition of two computational basis states $\ket{0}$ and $\ket{1}$, \ie{} \looseness=-1
\begin{equation*}
    \ket{\psi} = \alpha_0 \ket{0} + \alpha_1 \ket{1}, \ \alpha_0, \alpha_1 \in \mathbb{C} \ \text{and} \ \vert \alpha_0 \vert^2 + \vert \alpha_1 \vert^2 = 1 \,.
\end{equation*}
The notation $\ket{\cdot}$ (referred to as ket) is standard in quantum mechanics and is called Dirac notation. It holds that
\begin{equation}
    \ket{0}= \begin{bmatrix} 1 \\ 0  \end{bmatrix} \ \text{and} \ \ket{1}= \begin{bmatrix} 0 \\ 1  \end{bmatrix} ,
    \nonumber
\end{equation}
and can be thought of as analogous to the classical $0$ and $1$ states of a bit. Amplitudes $\boldsymbol \alpha$ are not directly accessible. Measurement of $\ket{\psi}$ results in states $\ket{0}$ and $\ket{1}$ with a probability $\vert \alpha_0 \vert^2$ and $\vert \alpha_1 \vert^2$, respectively. For instance, measuring $\ket{\psi} = \frac{i}{\sqrt{2}}\ket{0} - \frac{1}{\sqrt{2}}\ket{1}$ results in $\ket{0}$ with a probability $\vert \frac{i}{\sqrt{2}}\vert^2=\frac{1}{2}$ and in $\ket{1}$ with a probability $\vert \frac{1}{\sqrt{2}}\vert^2=\frac{1}{2}$.\footnote{It is common practice in electrical engineering to denote the imaginary unit by $j$ instead of $i$ to avoid confusion with the standard symbol for electric current. In this paper, however, we adopt the latter convention to remain consistent with the notation typically used in quantum information literature. When $i$ is used in this paper to denote current---as is standard in power electronics---this should be clear from the context.}

Systems of multiple qubits are obtained by combining the states of individual qubits through tensor products. For example, the combined state of a two-qubit system
\begin{equation*}
    \ket{\psi_1} = \alpha_0 \ket{0} + \alpha_1 \ket{1} \textrm{ and } \ket{\psi_2} = \alpha_0^\prime \ket{0} + \alpha_1^\prime \ket{1}
\end{equation*}
is 
\begin{align*}
    \ket{\psi_1} \otimes \ket{\psi_2} &= \ket{\psi_1 \psi_2}
    \\ 
    &= \alpha_0 \alpha_0^\prime \ket{00} {+} \alpha_0\alpha_1^\prime\ket{01} {+} \alpha_1 \alpha_0^\prime \ket{10} {+} \alpha_1 \alpha_1^\prime \ket{11}
    \\
    &= \alpha_{00} \ket{00}+\alpha_{01}\ket{01}+\alpha_{10} \ket{10}+\alpha_{11} \ket{11} .
\end{align*}
In $n$-qubit systems, the probability of measuring $\mathbf{x}$ with $\mathbf{x} \in \{0, 1 \}^n$ (also referred to as a bitstring) is $\vert \alpha_x \vert^2$, where $\alpha_x$ is the amplitude of the computational basis state $\ket{\mathbf{x}}$.

\subsubsection{Evolution of Quantum States}

In the gate model of quantum computation, the evolution of quantum states is realized by applying a unitary matrix (referred to as gate) $U$ to the state of the closed $n$-qubit system, such that $U^\dagger U = I$, where $U^\dagger$ is the conjugate transpose (adjoint) of $U$, and $I$ the identity matrix. Applying $U$ to a quantum state $\ket{\psi_1}$ transforms it into a new state $\ket{\psi_2}=U\ket{\psi_1}$. For example, the Hadamard gate that puts a single qubit in superposition is
\begin{equation*}
    H = \frac{1}{\sqrt{2}} \begin{bmatrix} 1 & 1 \\ 1 & -1 \end{bmatrix} .
\end{equation*}
Applying the Hadamard gate to $\ket{0}$ implies $H \ket{0}$ and results in a new state $\frac{1}{\sqrt{2}}\ket{0}+\frac{1}{\sqrt{2}}\ket{1}$. Other single-qubit gates that are relevant for this discussion are the Pauli Z and X matrices defined as 
\begin{equation*}
    \sigma^z = \begin{bmatrix} 1 & 0 \\ 0 & -1 \end{bmatrix}  \textrm{ and } \sigma^x = \begin{bmatrix} 0 & 1 \\ 1 & 0 \end{bmatrix} ,
\end{equation*}
respectively. Note that in a system of $n$ qubits, when single-qubit gates are independently applied to different qubits, the combined effect of the gates on the system is given by the tensor product of the individual gate operators.

Gates can also operate on more than one qubit. For instance, the controlled-NOT gate ($CX$), defined as
\begin{equation*}
    CX = \begin{bmatrix} 1 & 0 & 0 & 0 \\
                           0 & 1 & 0 & 0 \\
                           0 & 0 & 0 & 1 \\
                           0 & 0 & 1 & 0
            \end{bmatrix} ,
\end{equation*}
acts on a pair of qubits, namely a control and a target qubit. It flips the state of the target qubit, \ie{} it applies a NOT operation, if (and only if) the control qubit is $\ket{1}$. For example, applying the $CX$ gate to the state $\ket{\psi} = \frac{1}{\sqrt{2}}(\ket{00}+\ket{10})$ results in the new state 
\begin{align*}
  CX\ket{\psi} &= \frac{1}{\sqrt{2}} CX(\ket{00}+\ket{10}) = \frac{1}{\sqrt{2}}(CX\ket{00}+CX\ket{10})
  \\
  &= \frac{1}{\sqrt{2}}(\ket{00}+\ket{11})   
\end{align*}
because $\ket{00}$ remains unchanged, while the target qubit in $\ket{10}$ is flipped (since the control qubit is $\ket{1}$), turning it into $\ket{11}$.

Single- and multi-qubit gates can also be parametrized. For example, the $R_X$, $R_Z$ and $R_{ZZ}$ gates take an angle $\theta$ as input and are defined as 
\begin{equation*}
    R_X(\theta)= \begin{bmatrix} \cos(\frac{\theta}{2}) & -i\sin(\frac{\theta}{2}) \\ -i\sin(\frac{\theta}{2}) & \cos(\frac{\theta}{2}) \end{bmatrix} ,
\end{equation*}
\begin{equation*}
    R_Z(\theta)= \begin{bmatrix} e^{-i \frac{\theta}{2}} & 0 \\ 
                                0 & e^{i \frac{\theta}{2}} \end{bmatrix} ,
\end{equation*}

and 
\begin{equation*}
    R_{ZZ}(\theta) = \begin{bmatrix} e^{-i\frac{\theta}{2}} & 0 & 0 &0 \\ 
                               0 & e^{i\frac{\theta}{2}} & 0 & 0 \\
                               0 & 0 & e^{i\frac{\theta}{2}} & 0 \\
                               0 & 0 & 0 & e^{-i\frac{\theta}{2}}\end{bmatrix} .
\end{equation*}

\subsubsection{Projective Measurement}

Consider a unitary matrix $U$ and an orthonormal set of its eigenvectors $\ket{j}$ with corresponding eigenvalues $\lambda_j$. Its diagonal representation is given by $D = \sum_j \lambda_j \ket{j} \bra{j}$, where its term $\ket{j} \bra{j}$ is the outer product of the eigenvector $\ket{j}$. The matrix $P=\sum_j \ket{j}\bra{j}$ is called a projector onto the subspace spanned by the eigenvectors of $U$. The notation $\bra{\psi}$ (called a bra) indicates the dual vector of $\ket{\psi},$ \ie{} its conjugate transpose. Given two states $\ket{\psi_1}$ and $\ket{\psi_2}$, the expression $\braket{\psi_2 \vert \psi_1}$ represents their inner product and $\ket{\psi_1}\bra{\psi_2}$ their outer product.

In quantum computing, a measurement is described by an observable, which is a Hermitian operator $O$ that satisfies $O = O^\dagger$. Two important properties of Hermitian operators are that they have only real eigenvalues and that they can be diagonalized. The possible measurement outcomes correspond to the eigenvalues $\lambda_j$ of the observable. The observable can be written as $O = \sum_j \lambda_j P_j$, where each $P_j$ is a projector onto the eigenspace associated with eigenvalue $\lambda_j$. Given a quantum state $\ket{\psi}$, the probability of measuring $\lambda_j$ is $\pi_j = \braket{\psi\vert P_j\vert \psi}$, and the expected value of projective measurements is $\braket{O} = \braket{\psi \vert O \vert\psi}$.

It is straightforward to verify that $\sigma^z$ is a Hermitian operator with eigenvalues $+1$ and $-1$, and corresponding (orthonormal) eigenvectors $\ket{0}$ and $\ket{1}$, respectively. Based on the above, $\sigma^z$ can be written in its spectral decomposition as
\begin{equation*}
    \sigma^z = 1 \cdot \ket{0}\bra{0} - 1 \cdot \ket{1}\bra{1} \,.
\end{equation*}
Given the state $\ket{\psi}=\frac{1}{\sqrt{2}} \ket{0} + \frac{1}{\sqrt{2}} \ket{1}$, the probability of measuring $+1$ is 
\begin{equation*}
 \pi_{+1} = \bra{\psi}\ket{0} \bra{0}\ket{\psi} = |\braket{0 \vert \psi}|^2 = \frac{1}{2} \,,
\end{equation*}
and the probability of measuring $-1$ is
\begin{equation*}
 \pi_{-1} = \bra{\psi}\ket{1} \bra{1}\ket{\psi} = |\braket{1 \vert \psi}|^2 = \frac{1}{2} \,.
\end{equation*}
Similarly, the expected value of projective measurements of $\ket{\psi}$ is
\begin{equation*}
    \braket{\sigma^z} = \braket{\psi \vert \sigma^z \vert \psi} = \bra{\psi}(\ket{0}\bra{0} - \ket{1}\bra{1})\ket{\psi} = 0 \,.
\end{equation*}
It is important to note that, after a measurement, the state of the qubit collapses to the measured computational basis state. In the case of an $n$-qubit system, this means the post-measurement state is $\ket{\mathbf{x}}$ for some $\mathbf{x} \in \{0, 1 \}^n$.

\subsubsection{Quantum Circuits}

\begin{figure}[t!]
    \centering
    \includegraphics[width=0.9\linewidth]{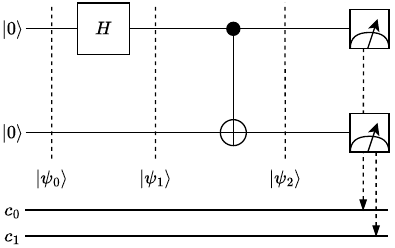}
    \caption{A quantum circuit that prepares and measures the state $\frac{1}{\sqrt{2}}(\ket{00}+\ket{11})$, which is also one of the Bell states.}
    \label{fig:example_circuit}
    \vspace{\figSpace}
\end{figure}

A quantum algorithm is a sequence of gates applied to a system of qubits, often called a quantum circuit. An example is shown in Fig.~\ref{fig:example_circuit} (note that the visual presentation of quantum circuits is not strictly standardized). In this figure, each thin wire corresponds to a qubit, and each bold wire to a classical register. The gates are applied from the left to right and the wires from top to bottom.
The example represents a system of two qubits, each initialized to $\ket{0}$, \ie{} the initial state of the system is $\ket{\psi_0}=\ket{00}$. Next, a Hadamard gate is applied to the first qubit, whereas no operator acts on the second qubit (which is equivalent to applying the identity matrix). This yields the state
\begin{equation*}
    \ket{\psi_1} = (H\otimes I) \ket{00} = \frac{1}{\sqrt{2}} (\ket{00} + \ket{10}) \,.
\end{equation*}
Following, a controlled-NOT ($CX$) gate is applied with the first qubit being the control and the second the target. This produces the state
\begin{equation*}
    \ket{\psi_2} = CX \ket{\psi_1} = \frac{1}{\sqrt{2}} (\ket{00} + \ket{11}) \,.
\end{equation*}
At this point, a projective measurement of both qubits on the computational basis is performed. The projectors are $\ket{00}\bra{00}$, $\ket{01}\bra{01}$, $\ket{10}\bra{10}$, and $\ket{11}\bra{11}$. It can be verified that the only possible outcomes upon measurement are the bitstrings $00$ and $11$, each occurring with equal probability. These outcomes are recorded in the classical registers $c_1$ and $c_2$, respectively. \looseness=-1

\section{QAOA: Didactic Example}
\label{app:qaoa-example}

A step-by-step example of applying QAOA that was described in Section \ref{sec:qaoa_theory} is presented hereafter. We seek the ground state of the following QUBO problem with four decision variables:
\begin{mini*}
{\mathbf{x} \in \left\{0, 1\right\}^4}
{\!\!\! f(\mathbf{x}) \!=\! -3 x_0 - 3 x_3 + 2 x_0 x_1 + 2 x_1 x_2 + 2 x_2 x_3 .}
{\label{example_QUBO}}{}
\end{mini*}
There are $2^4$ possible solutions. It can be easily verified that this problem attains its unique minimum value $f(\mathbf{x^*})=-6$ at the solution $\mathbf{x}^*=(1, 0, 0, 1)$. Solving this problem by uniformly sampling from the possible bitstrings has a $6.25$\% probability of successfully finding the ground state. 

\begin{figure}[t]
    \centering
    \includegraphics[width=1\linewidth]{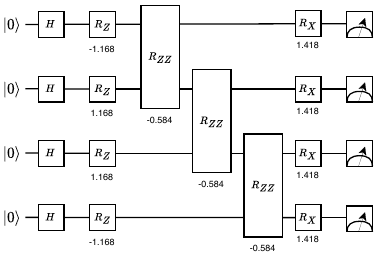}
    \caption{Optimized QAOA ansatz for the didactic example for $p=1$. Numerical values below the gate symbols indicate the optimized parameters.}
    \label{fig:didactic_example_circuit}
\end{figure}

\begin{figure}[t]
    \centering
    \includegraphics[width=1\linewidth]{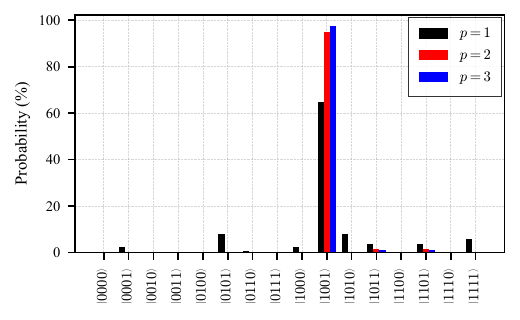}
    \caption{Post-measurement computational state probabilities for the didactic example for different values of $p$. The expected energy after training the QAOA ansatz is $\Braket{\mathcal{H}^C}_1=-4.77$, $\Braket{\mathcal{H}^C}_2=-5.84$, and  $\Braket{\mathcal{H}^C}_3=-5.93$.\looseness=-1}
    \label{fig:didactic_example_probs}
\end{figure}

To solve the problem of interest using QAOA, the first step is to convert the QUBO problem into an Ising spin model:
\begin{mini*}
{\mathbf{z} \in \left\{-1, +1\right\}^4}
{f^\prime(\mathbf{z}) ~&= -\frac{3}{2} + z_0 - z_1 - z_2 + z_3}{}{}
\breakObjective{+ \frac{1}{2} z_0 z_1 + \frac{1}{2} z_1 z_2 + \frac{1}{2} z_2 z_3} \,.
{\label{example_ising}}{}
\end{mini*}
Promoting each spin variable to a Pauli Z matrix, the associated cost Hamiltonian is
\begin{equation*}
\begin{split}
    \mathcal{H}^C &= -\frac{3}{2} I_0 + \sigma_0^z - \sigma_1^z - \sigma_2^z + \sigma_3^z  \\&+\frac{1}{2} \sigma_0^z \otimes \sigma_1^z +\frac{1}{2} \sigma_1^z \otimes \sigma_2^z +\frac{1}{2} \sigma_2^z \otimes \sigma_3^z \,.
\end{split} 
\end{equation*}
We select the standard mixer Hamiltonian $\mathcal{H}^M=\sum_{i=0}^{3} \sigma_i^x$, where $\sigma_i^x$ is the Pauli X matrix acting on qubit $i$. Therefore, for a given depth $p \geq 1$, the QAOA unitary for this problem is defined as
\begin{equation*}
    U(\beta, \gamma) = \left[\prod_{j=1}^{p} e^{-i\beta_j \mathcal{H}^M} e^{-i\gamma_j \mathcal{H}^C} \right] H^{\otimes 4} \,,
    \label{QAOA_example_problem}
\end{equation*}
where $i^2=-1$, while $\boldsymbol \beta$, $\boldsymbol \gamma \in \mathbb{R}^p$ are the $2p$ variational parameters to be estimated, and $H$ is the Hadamard unitary. The optimized QAOA ansatz implemented using $R_X$, $R_Z$ and $R_{ZZ}$ gates for $p=1$ is shown in Fig.~\ref{fig:didactic_example_circuit}. The probability of observing different states is shown in Fig.~\ref{fig:didactic_example_probs} for different values of $p$. We observe that the optimal solution of the optimization problem $\ket{1001}$ can be obtained with high probability ($>95$\%) already for $p=2$, whereas $\braket{\mathcal{H}^C}$ approaches $f(\mathbf{x^*})=-6$. \looseness=-1

\bibliographystyle{ieeetran}
\bibliography{bibliography}


\end{document}